\documentclass[aps,prb,twocolumn,showpacs,nofootinbib]{revtex4-1}
\usepackage{graphicx}
\usepackage[caption=false]{subfig}
\usepackage{float}
\usepackage{longtable}
\usepackage{bm}
\usepackage{siunitx}
\usepackage{xfrac}
\usepackage{amssymb,mathtools,mathrsfs,bbold,dsfont}
\usepackage{amstext,braket,amsmath,tensor}  
\usepackage{array} 
\usepackage{siunitx}
\usepackage[normalem]{ulem}
\usepackage{empheq}
\usepackage{multirow}
\usepackage{amsthm,etoolbox}
\usepackage{cleveref}

\newcommand{\bq}{\boldsymbol{q}}

\newcommand{\Vcal}{\mathcal{V}}

\newcommand{\Rcal}{\mathcal{R}}

\newcommand{\Hcal}{\mathcal{H}}

\newcommand{\Zcal}{\mathcal{Z}}

\newcommand{\bPi}{\boldsymbol{\Pi}}
\newcommand{\bM}{\boldsymbol{M}}

\newcommand{\bG}{\boldsymbol{G}}

\newcommand{\bR}{\boldsymbol{R}}

\newcommand{\bl}{\boldsymbol{l}}

\newcommand{\bD}{\boldsymbol{D}}

\newcommand{\bRcal}{\boldsymbol{\Rcal}}

\newcommand{\bLambda}{\boldsymbol{\Lambda}}

\newcommand{\eq}{\scriptscriptstyle{\text{eq}}}

\renewcommand{\Im}{\text{Im}}
\renewcommand{\Re}{\text{Re}}

\newcommand{\avg}[1]{\left\langle{#1}\right\rangle}
\newcommand{\Tr}[1]{\textup{Tr}\left[{#1}\right]}

\newcommand{\angs}{\textup{\AA}}

\newcommand{\bPhi}{\boldsymbol{\Phi}}

\newcommand{\Bavg}[1]{\Bigl\langle #1 \Bigr\rangle}

\newcommand{\ssn}{\scriptscriptstyle{(n)}}

\newcommand{\ssF}{\scriptscriptstyle{(F)}}
\newcommand{\sst}{\scriptscriptstyle{(3)}}
\newcommand{\ssf}{\scriptscriptstyle{(4)}}

\DeclareMathAlphabet{\mathbbmsl}{U}{bbm}{m}{sl}

\newcommand{\ssz}{\scriptscriptstyle{(0)}}
\DeclareSymbolFont{bbold}{U}{bbold}{m}{n}
\DeclareSymbolFontAlphabet{\mathbbold}{bbold}

\newcommand{\ssS}{\scriptscriptstyle{(S)}}

\newcommand{\ssB}{\scriptscriptstyle{(B)}}

\newcommand{\Rcaleq}{\Rcal_{\scriptscriptstyle{\text{eq}}}}
\newcommand{\bRcalhs}{\bRcal_{\scriptscriptstyle{\text{hs}}}}
\newcommand{\bRcaleq}{\bRcal_{\scriptscriptstyle{\text{eq}}}}
\newcommand{\B}{\scriptscriptstyle{\text{B}}}

\newcommand{\Hsyst}{\text{H}_3\text{S}}
\newcommand{\Dsyst}{\text{D}_3\text{S}}

\newcommand{\bcc}{\scriptscriptstyle{\text{bcc}}}
\newcommand{\ssZPE}{\scriptscriptstyle{\text{ZPE}}}

\begin{document}

\title{High-pressure phase diagram of hydrogen and deuterium sulfides from first principles: structural and vibrational properties including quantum and anharmonic effects.}
\author{Raffaello Bianco$^{1,2}$}
\email{raffaello.bianco@roma1.infn.it}
\author{Ion Errea$^{3,4}$}
\author{Matteo Calandra$^{5}$}
\author{Francesco Mauri$^{1,2}$}

\affiliation{$^{1}$ Dipartimento di Fisica, Universit\`a di Roma La Sapienza, 
Piazzale Aldo Moro 5, I-00185 Roma, Italy}
\affiliation{$^{2}$ Graphene Labs, Fondazione Istituto Italiano di Tecnologia, Via Morego, I-16163 Genova, Italy}
\affiliation{$^3$Fisika Aplikatua 1 Saila, Bilboko Ingeniaritza Eskola, University of the Basque Country (UPV/EHU),
Rafael Moreno ``Pitxitxi'' Pasealekua 3, 48013 Bilbao, Basque Country, Spain}
\affiliation{$^4$Donostia International Physics Center (DIPC),
            Manuel de Lardizabal pasealekua 4, 20018 Donostia-San Sebasti\'an,
	    Basque Country, Spain}
\affiliation{$^5${Sorbonne Universit\'e, CNRS, Institut des
  Nanosciences de Paris, UMR7588, F-75252, Paris, France}}

\begin{abstract}
{We study the structural and vibrational properties of the high-temperature superconducting
sulfur trihydride and trideuteride in the high-pressure $Im\bar{3}m$ and $R3m$ phases 
by first-principles density-functional-theory calculations. 
On lowering pressure, the rhombohedral transition $Im\bar{3}m \rightarrow R3m$ is expected,
with hydrogen bond desymmetrization and occurrence of trigonal lattice distortion.
With both PBE and BLYP exchange-correlation functional, 
in hydrostatic conditions we find that, contrary to what suggested in some recent experiments,
if the rhombohedral distortion exists it affects mainly the hydrogen-bonds, whereas 
the resulting cell distortion is minimal. We estimate that the occurrence of a 
stress anisotropy of approximately $10\%$ could explain this discrepancy.
Assuming hydrostatic conditions, we calculate the critical pressure at which the 
rhombohedral transition occurs. Quantum and anharmonic effects, which are relevant in this system, 
are included at nonperturbative level with the stochastic self-consistent harmonic approximation.
Within this approach, we determine the transition pressure by calculating the free energy Hessian, a method 
that allows to estimate the critical pressure with much higher precision (and much lower computational cost) 
compared with the free energy `finite-difference' approach previously used.
Using PBE and BLYP, we find that quantum anharmonic effects are responsible for a strong reduction of the critical 
pressure with respect to the one obtained with the classical harmonic approach. Interestingly,
for the two functionals, even if the transition pressures at classical harmonic level differ by 83~GPa, 
the transition pressures including quantum anharmonic effects differ only by 23~GPa.
Moreover, we observe a prominent isotope effect, as we estimate higher pressure transition 
for D${}_3$S than for H${}_3$S. 
Finally, within the stochastic self-consistent harmonic approximation, with PBE 
we calculate the anharmonic phonon spectral functions in the
$Im\bar{3}m$ phase. The strong anharmonicity of the system is confirmed by
the occurrence of very large anharmonic broadenings leading to complex non-Lorentzian 
line shapes. Generally, for the high-energy hydrogen bond-stretching modes, the anharmonic
phonon broadening is of the same magnitude of the electron-phonon one. However, for the vibrational spectra at zone center, 
accessible e.g. by infrared spectroscopy, the broadenings are very small (linewidth at most around 2~meV) 
and anharmonic phonon quasiparticles are well defined.}
\end{abstract}

\maketitle

\newtoggle{draft}
\togglefalse{draft}
\section{Introduction}
\label{sec:Introduction}
High-temperature superconductivity in compressed hydrogen sulfide has been recently reported, 
with a record critical temperature $T_c$ above 200 K at $P\simeq 150$ GPa~\cite{Drozdov2015}. 
This discovery, which had been anticipated by \textit{ab initio} calculations~\cite{Duan2014},
{stimulated} a number of experimental~\cite{Einaga2016,PhysRevB.93.174105,
PhysRevB.95.140101,Capitani2017,PhysRevB.95.020104,PhysRevB.93.020103} and theoretical 
studies~\cite{PhysRevB.93.020103,PhysRevLett.114.157004,Errea2016,PhysRevLett.117.075503,PhysRevB.91.224513,
Flores-Livas2016,PhysRevB.91.180502,0295-5075-112-3-37001,HIRSCH201545,ANIE:ANIE201511347,ANGE:ANGE201704364}
aiming at characterizing this fascinating compound. {Even if it seems
established that the electron-phonon mechanism and strong anharmonic effects
are the key factors in determining the high superconducting critical
temperature, the pressure phase diagram of the H-S system is still
controversial, mainly due to conflicting experimental results}.
 
The overall consensus is that H${}_2$S, 
the only stable stoichiometry formed by hydrogen and sulfur at ambient conditions, 
decomposes at high pressures giving rise to several H-S compounds, 
typically sulfur and hydrides with a larger hydrogen content. 
Indeed, at high pressure H${}_2$S  is a potential superconductor, but with a relatively low $T_c$ (around 80 K~\cite{doi:10.1063/1.4874158}),
and is metastable. The experimental observation of a rapid increase of $T_c$ on pressure above approximately 150 GPa~\cite{Drozdov2015}
has been attributed to the decomposition of H${}_2$S and the formation
of H${}_3$S through the $3\text{H}_{2}\text{S}\rightarrow2\text{H}_{3}\text{S}+\text{S}$ decomposition.
Theoretical calculations show that the decomposition of H${}_2$S
into $\text{H}_{3}\text{S}+\text{S}$ is energetically favored already at low pressures
and that a 200 K superconductivity is consistent with the H${}_3$S 
stoichiometry assuming the conventional electron-phonon pairing with
anharmonic phonons~\cite{PhysRevB.93.020103,PhysRevLett.114.157004,
Errea2016,PhysRevLett.117.075503,PhysRevB.91.224513,
Flores-Livas2016,PhysRevB.91.180502}. 
Intermediate `Magn\'eli' phases emerging between the H${}_2$S to  H${}_3$S decomposition
seem to explain the rapid soar of $T_c$ on pressure loading~\cite{PhysRevLett.117.075503}.

First-principles structural prediction simulations within density functional theory (DFT)
and the Perdew-Burke-Ernzerhof (PBE) 
parametrization for the exchange-correlation functional yield a phase diagram for H${}_3$S
consisting of a $Cccm \to R3m \to Im\bar{3}m$ sequence~\cite{Duan2014}.
The $Cccm \to R3m$ transition is predicted at  approximately 110 GPa
and the transition to the $Im\bar{3}m$ at 180 GPa. Other predictions
give the same phase diagram above 110 GPa, but a $C2/c$ phase below~\cite{PhysRevB.93.020103}.
H${}_3$S  becomes metallic after the $Cccm \to R3m$ or  $C2/c \to R3m$ transition, once
the H${}_2$S and H$_2$ units present in both $Cccm$ and $C2/c$ structures are broken. 
In the $Im\bar{3}m$ phase the sulfur atoms of $\Hsyst$
are arranged in a body-centered cubic (bcc) lattice and each hydrogen atom resides halfway between 
two adjacent sulfur atoms, 
forming symmetric straight S-H-S bonds. 
The $R3m$ phase is obtained from $Im\bar{3}m$ with a displacive transition that 
reduces the cubic symmetry with a rhombohedral distortion.
This distortion, however, is estimated~\cite{Errea2016} to have a small impact on the position 
of the sulfur atoms, which remain essentially unaffected on a bcc lattice, 
whereas the hydrogen atoms are shifted from 
symmetric positions in the S-H-S bonds (see Fig.~\ref{fig:competing_structures}). 

Early x-ray diffraction (XRD) measurements~\cite{Einaga2016} confirmed the 
bcc arrangement of the S atoms at high pressure, but could not distinguish
between the $R3m$ and  $Im\bar{3}m$ phases, as H atoms are very weak scatters 
and their position cannot be determined.
However, the pronounced kink observed at $150$ GPa in the pressure dependence of $T_c$
could signal the occurrence of a phase transition (widely reputed of second-order), possibly
the $R3m \to Im\bar{3}m$ transition~\cite{Einaga2016}. 
Two more recent experimental works that use as direct reactants H${}_2$ and S in order to synthesize 
perfect crystallized samples with $\Hsyst$ stoichiometry (as opposed to the use of H${}_2$S as reactant in the previous experiments) 
show surprising, and somewhat contradicting, results (however, the disagreement might be 
due to the different pressure used for the synthesis of H${}_3$S and the large kinetic barriers that yield large metastability regions).
In the first place, Goncharov \emph{et al.}~\cite{PhysRevB.95.140101} confirm the pressure induced destabilization 
of H${}_2$S in favor of H${}_3$S and the theoretically predicted $Cccm$-$R3m$-$Im\bar{3}m$ structural sequence.
In these experiments, the cubic $Im\bar{3}m$ was directly synthesized at large pressure, and subsequent pressure release 
led to the appearance of a rhombohedral distortion compatible with the $R3m$ phase. The 
$R3m$ remained metastable down to 70 GPa, where it transformed upon annealing to the $Cccm$.
However, contrary to what expected, the trigonal lattice distortion observed around 110 GPa was very large compared to the
calculations~\cite{Duan2014,Errea2016}, although the authors admit the possibility that it could be a
consequence of nonhydrostatic pressure conditions. Secondly, Guigue \emph{et al.}~\cite{PhysRevB.95.020104} provide evidence supporting that $\Hsyst$ is
the stoichiometry of the sulfur hydride observed at high pressures (above 75 GPa) 
but, quite surprisingly, they only observed the orthorhombic $Cccm$ phase up to 160 GPa, 
thus contradicting the expected $Cccm$-$R3m$-$Im\bar{3}m$ sequence
and suggesting that a metallic transition to the $R3m$ or $Im\bar{3}m$ could happen at higher pressures or 
that the system can be trapped in a $Cccm$ metastable phase.
Several other hypothesis are also proposed by the authors in order to account for this result, essentially based
on the possibility that the experimental high-pressure  XRD patterns, usually ascribed to $\Hsyst$, 
belong to the diffraction pattern of a compound with another stoichiometry.

In this paper we present a detailed theoretical first-principle analysis 
of the structural and vibrational properties of high pressure $\Hsyst$ 
and $\Dsyst$ in the $Im\bar{3}m$ and $R3m$ phases. All the force-energy and linear response harmonic calculations have been performed
with the density-functional theory (DFT) method, within the generalized gradient approximation 
(GGA) for the exchange-correlation functional. We mainly utilize the Perdew-Burke-Ernzerhof (PBE) 
parametrization for GGA (see App.~\ref{sec:Calculation_details}). The choice of PBE is motivated by the better agreement between the 
theoretical and experimental equation of state, if compared with other functionals (see~Ref.~\citenum{Errea2016}), 
coherently with the fact that PBE is known to correctly reproduce the compression
curve of hydrides~\cite{PhysRevLett.113.265504,PhysRevB.93.224104}. However,
in order to estimate the impact on the results of the approximations inherent DFT, in several cases
we also present results obtained with the Becke-Lee-Yang-Parr (BLYP) parametrization of GGA.

With harmonic calculations including zero point motion
we analyze if the occurrence of a $Im\bar{3}m\rightarrow R3m$ transition can effectively have a strong impact
not only on the H atoms positions, but also on the bcc arrangement of the S atoms.
We also analyze the role that nonhydrostaticity can play.
%
We then study $\Hsyst$ at high pressure in the conventional hydrostatic setting and,
in order to  fully include anharmonic and quantum effects, we utilize the stochastic self-consistent harmonic approximation
(SSCHA)\cite{PhysRevB.89.064302,PhysRevLett.111.177002}. We apply the SSCHA extension presented in Ref.~\citenum{PhysRevB.89.064302}
to evaluate, with the free energy Hessian, the transition pressure $P_c$ between the $Im\bar{3}m$ and $R3m$ phases.
In Ref.~\citenum{Errea2016} the SSCHA was already applied to evaluate the critical pressure of the $Im\bar{3}m \rightarrow R3m$ transition, 
but explicitly estimating the free energy difference between the two phases. This direct approach, however, is computational demanding 
as it requires, in principle, separate SSCHA calculations for several average atomic positions along selected distortion paths. Moreover, 
SSCHA calculations in low-symmetry phases are more delicate than SSCHA calculations in the high-symmetry phase, because 
at low symmetry it is necessary to consider a large number of configurations to reduce
the statistical uncertainly and obtain reliable converged estimations of the free energy. 
Indeed, in Ref.~\citenum{Errea2016} explicit SSCHA calculations in the low-symmetry phases where performed only for two volumes 
(more details in Sec.~\ref{sec:Critical_pressure_of_the_trigonal_phase_transition}).
On the contrary, with the free energy Hessian method adopted in this work, for each volume we have complete and direct access to the lattice instability of the system
with workload almost comparable with the one of a single SSCHA free energy calculation in the high-symmetry phase. 
This allows to calculate the critical pressure $P_c$ with much higher precision than any `finite-difference' approach. 
Finally, in order to facilitate the comparison with  experimental results, we apply the SSCHA dynamical extension of Ref.~\citenum{PhysRevB.89.064302} to calculate
the anharmonic phonon spectral functions and dispersions.

The paper is structured as follows.
In Sec.~\ref{sec:Effect_of_nonhydrostatic_pressure} we analyze the extent of the trigonal lattice distortion expected in the 
 $Im\bar{3}m\rightarrow R3m$ rhombohedral transition, and  we estimate the effect of nonhydrostatic pressure.
In Sec.~\ref{sec:Critical_pressure_of_the_trigonal_phase_transition}, under hydrostatic pressure conditions, 
we compute the critical pressure $P_c$ of the $Im\bar{3}m\rightarrow R3m$
transition for $\Hsyst$ and $\Dsyst$,
within the SSCHA free energy Hessian method~\cite{PhysRevB.89.064302}.
In Sec.~\ref{sec:Spectrum_and_dispersion_of_anharmonic_phonons_in_the_high-pressure_bcc_phase} we use 
the SSCHA \emph{dynamical ansatz}~\cite{PhysRevB.89.064302} to calculate the anharmonic phonon spectral properties of 
$\Hsyst$ ($\Dsyst$) in the high-pressure $Im\bar{3}m$ phase.
In Sec.~\ref{sec:Conclusions}, we summarize our results and draw some final conclusions. The
paper is completed with two appendices. 
In App.~\ref{sec:Theoretical_Method} we summarize the theoretical method used.
In App.~\ref{sec:Calculation_details} we describe the computational details of the calculations 
and the results of the convergence tests.

\section{Extent of the trigonal lattice distortion and effect of nonhydrostatic pressure}
\label{sec:Effect_of_nonhydrostatic_pressure}
\begin{figure}[t!]
\centering
\includegraphics[width=\columnwidth]{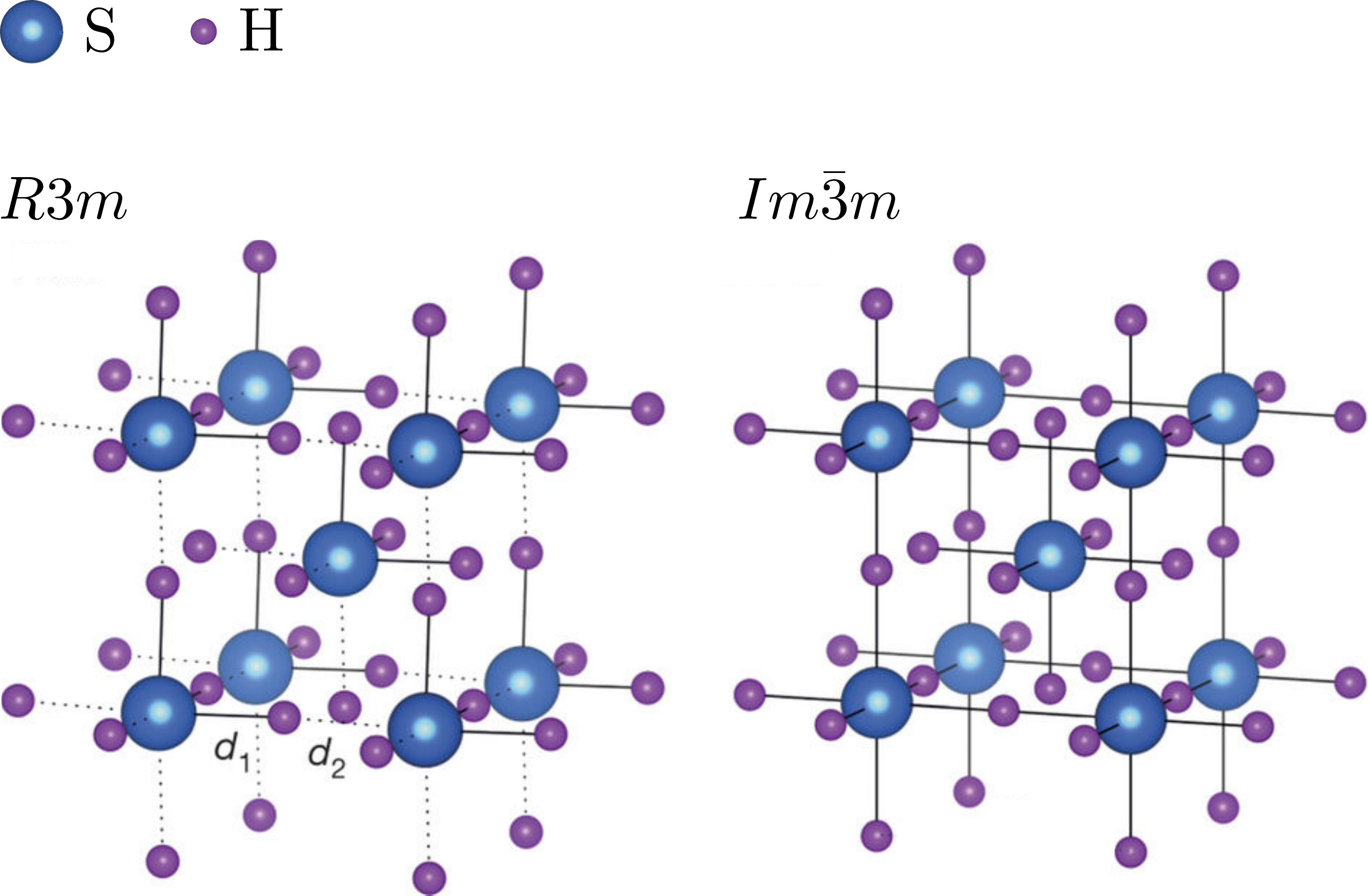}
\caption{(Color online) Crystal structure in the conventional bcc cell of the $R3m$ (left) and $Im\bar{3}m$ (right)
phases of $\Hsyst$. In the $R3m$ structure, the H$-$S covalent bond of length $d_1$ is marked with a 
solid line and the longer H$\cdot\!\!\cdot\!\!\cdot$S hydrogen
bond of length $d_2$ with a dotted line. In the $Im\bar{3}m$ phase $d_1=d_2$. 
Blue and pink spheres represents S and H atoms, respectively.}
\label{fig:competing_structures}
\end{figure}

At very high pressure hydrogen sulfide is expected to be in the trihydride stoichiometry with metallic structure $R3m$/$Im\bar{3}m$.
$R3m$ is a rhombohedral modification occurring to the $Im\bar{3}m$ bcc phase at lower pressures, with asymmetric H-S bonds. 
In Fig.~\ref{fig:competing_structures} we show these two structures. 
DFT calculations suggest that the rhombohedral distortion essentially involves only the position of the hydrogen 
atoms~\cite{Errea2016}, i.e. that the trigonal angle of the S atoms in the $R3m$ structure 
is very close to the bcc one, as in $Im\bar{3}m$.
However, recent measurements challenge this picture and report a 
large trigonal lattice distortion at pressures below 110 GPa 
(see Fig.~2 in Ref.~\citenum{PhysRevB.95.140101}), with a trigonal
angle $\vartheta$ {such that} $\cos \vartheta \simeq -0.324$.
The trigonal angle can be given in terms of the hexagonal $a_{\rm hex}$ and $c_{\rm hex}$ lattice parameters 
used in Ref.~\citenum{PhysRevB.95.140101}
as $\cos \vartheta = [2(c_{\rm hex}/a_{\rm hex})^2-3]/ [2(c_{\rm hex}/a_{\rm hex})^2+6]$. 
Ultrahigh pressure experiments for sulfur hydrides are performed with confinement in a 
diamond anvil cell (DAC).
The experimental setting is very delicate and special care is required 
in order to reach hydrostatic conditions.
Undesired nonhydrostatic pressure 
conditions can appear in experiments, which could be at the origin of such 
trigonal distortion~\cite{PhysRevB.95.140101}. Motivated by these considerations, we calculate within DFT-PBE
the total energy and the stress anisotropy of $\Hsyst$ in the orthorhombic phase $R3m$ 
as a function of the trigonal angle $\vartheta$ at fixed volume.
Here the stress anisotropy is measured by the difference $\sigma_{\parallel}-\sigma_{\bot}$, where 
$(\sigma_{\parallel},\sigma_{\bot},\sigma_{\bot})$ are the eigenvalues (principal stresses) 
of the stress tensor $\sigma_{ij}$ in the $R3m$ phase (notice that two of them are equal by symmetry).
For each cell configuration, the internal atomic positions have been relaxed until the forces acting on them are zero 
(thus $\vartheta$ is the only free parameter). 

\begin{figure*}[t!]
\centering
\includegraphics[width=\textwidth]{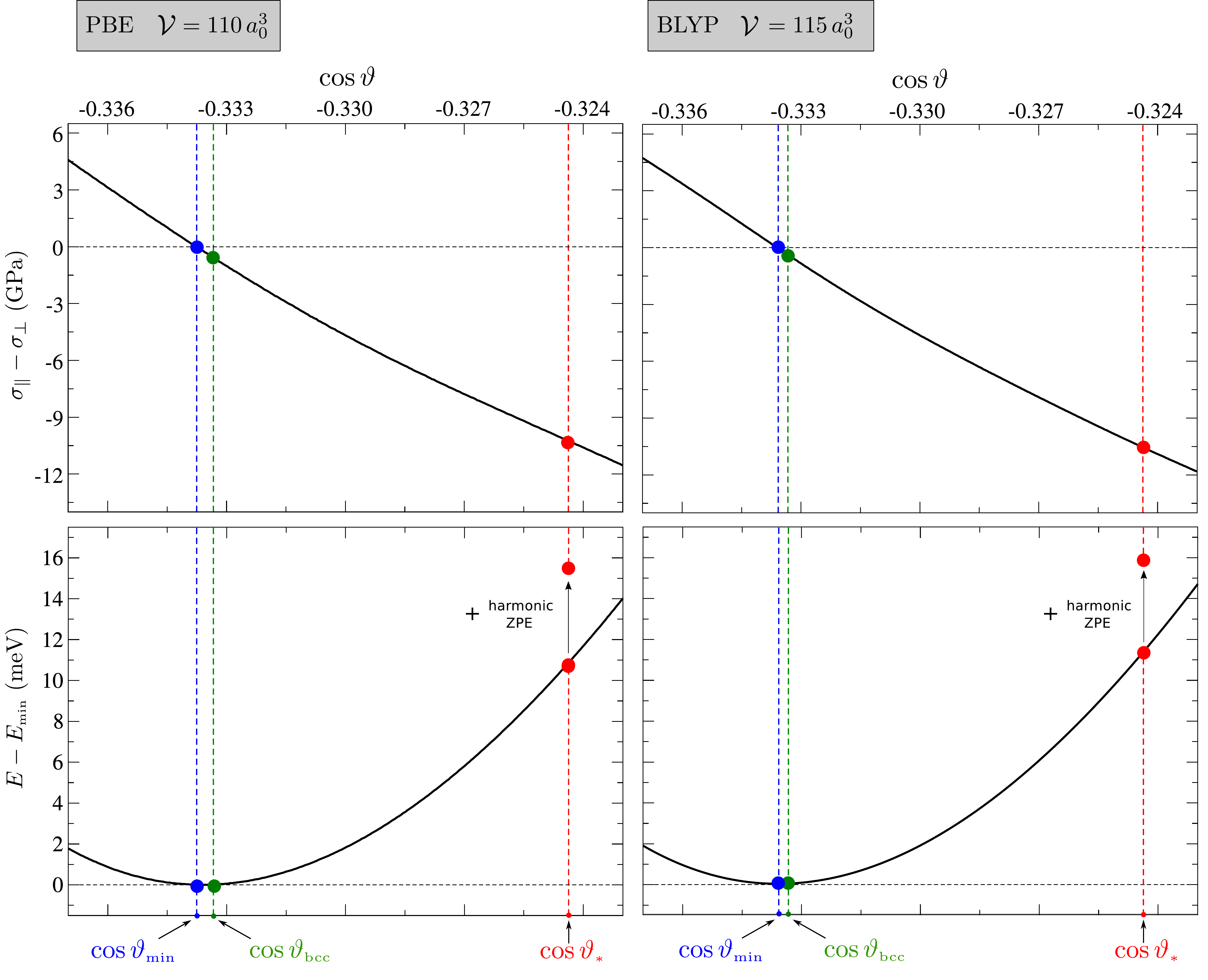}
\caption{(Color online) Lower panels: total energy (per unit cell) of H${}_3$S in the $R3m$ phase 
(with respect to the energy minimum) as a function of $\cos\vartheta$, where $\vartheta$ 
is the trigonal angle. The calculation including the ZPE at the harmonic level is also shown. 
Upper panels: stress anisotropy $\sigma_{\parallel}-\sigma_{\bot}$
as a function of $\cos\vartheta$ ($\sigma_{\parallel},\sigma_{\bot},\sigma_{\bot}$
are the eigenvalues of the stress tensor). The calculations are performed with PBE, in the left column, and with BLYP,
in the right column. The volume $\Vcal$ is kept fixed at $110\,a_0^3$,  in the first case, and at $115\,a_0^3$, 
in the second case ($a_0$ is the Bohr radius). In hydrostatic conditions, these volumes correspond to a theoretical pressure of $100$ GPa,
for the two functionals. 
For each angle, the internal positions are relaxed until the forces acting on atoms are less than $10^{-3}$ Ry/$a_0$. 
The angle $\vartheta_{\scriptscriptstyle{\text{bcc}}}$ corresponds to the body-centered cubic (bcc)
cell ($\cos\vartheta_{\scriptscriptstyle{\text{bcc}}}=-1/3$). 
The energy minimum is obtained with the angle $\cos\vartheta_{\scriptscriptstyle{\text{min}}}\simeq -0.33376$ in the PBE case, 
and $\cos\vartheta_{\scriptscriptstyle{\text{min}}}\simeq -0.33359$ in the BLYP case.
The $\vartheta_*$ angle measured in the experiment of Ref.~\citenum{PhysRevB.95.140101} at pressures below 110 GPa
($\cos\vartheta_*\simeq-0.324$) is marked for comparison.}
\label{fig:nonhydrostatic}
\end{figure*}

The volume is kept equal to 110 $a_0^3$, where $a_0=0.5292\,\angs$ is the Bohr radius.
With PBE, in hydrostatic conditions this volume corresponds to a (theoretical) pressure of 100 GPa. 
According to Fig.~2 in Ref.~\citenum{PhysRevB.95.140101}, at this pressure the system should be in equilibrium 
with a trigonal angle $\vartheta_*$ quite smaller than the bcc one $\vartheta_{\bcc}$ ($\cos\vartheta_{\scriptscriptstyle{\text{bcc}}}=-1/3$, 
$\cos\vartheta_*\simeq-0.324$). As we can see from Fig.~\ref{fig:nonhydrostatic}, 
this contradicts the results of our DFT calculations. 
At this volume, the most stable structure has isotropic stress and a trigonal angle 
$\vartheta_{\scriptscriptstyle{\text{min}}}$ which is very 
similar to the bcc one, only slightly larger ($\cos\vartheta_{\scriptscriptstyle{\text{min}}}\simeq -0.334$).
It is worthwhile to recall that, at this pressure, even with $\vartheta_{\bcc}$ the atoms 
relax in a configuration with $R3m$ space group that breaks the S--H--S bond symmetry, 
i.e. a configuration different from $Im\bar{3}m$.
In order to obtain the experimentally observed angle $\vartheta_*$ 
it is necessary to apply a significant stress anisotropy of $\sigma_{\parallel}-\sigma_{\bot} \sim 10$ GPa,
thus an anisotropy of around 10\%. This conclusion is strengthened if we include the vibrational contribution to the energy.
Indeed, the configuration with $\vartheta_*$ is even more unstable when the harmonic zero-point energy (ZPE) contribution 
is included (see Fig.~\ref{fig:nonhydrostatic}). The result is also robust with respect to the 
DFT exchange-correlation functional. Indeed, as shown in Fig.~\ref{fig:nonhydrostatic}, 
similar results are obtained with the BLYP parametrization for GGA. 
Notice that with this functional, in order to have theoretical pressure of 100 GPa (in hydrostatic conditions),
we keep the volume equal to 115 $a_0^3$. 

In conclusion, the results obtained in our calculations confirm the small distortion of the bcc 
lattice occurring in a $Im\bar{3}m\rightarrow R3m$ transition  and indicate that it is a nonhydrostatic pressure, 
occurred in the experiment, that could have caused the large rhombohedral distortion observed~\cite{PhysRevB.95.140101}.

\section{Critical pressure of the rhombohedral phase transition}
\label{sec:Critical_pressure_of_the_trigonal_phase_transition}
In the previous section we numerically confirmed that, as obtained in previous DFT calculations, in hydrostatic conditions 
if at high-pressure the transition $Im\bar{3}m \rightarrow R3m$ occurs, then
it has small impact on the positions of sulfur atoms, which essentially remain in a bcc configuration. However, 
while there is quite a wide theoretical consensus on this
aspect of the rhombohedral displacive transition of $\Hsyst$, the pressure transition $P_c$
at which it would occur is still uncertain. 
This is mainly due to two nontrivial aspects that have to be considered.
First, the quantum nature of the proton, which can crucially affect the structural and physical properties of hydrogen compounds.
That is particularly true for the transition $R3m\rightarrow Im\bar{3}m$, where a symmetrization of the hydrogen-bond 
occurs, similar to the quantum proton fluctuations leading to the symmetrization of hydrogen bonds in 
ice X~\cite{Goncharov218,Loubeyre1999}.
Moreover, since H has a very light mass and we deal with a phase transition, anharmonic effects become crucial 
as already determined in recent works~\cite{PhysRevLett.114.157004,Errea2016}. 
Motivated by these considerations, we apply the stochastic self-consistent harmonic approximation (SSCHA)~\cite{PhysRevB.89.064302}. 

The SSCHA is a variational method that allows to evaluate $F(\bRcal)$, 
the free energy of the system as a function of the average atomic position $\bRcal$ 
(with bold symbols we indicate quantities in component-free notation). 
$F(\bRcal)$ has the minimum in the equilibrium configuration $\bRcaleq$, which has to be compared with
the classical equilibrium position $\bRcaleq^{\ssz}$, the configuration for which $V(\bR)$, the potential energy felt by the nuclei 
(in the Born-Oppenheimer approximation), has the minimum. 

In the classical harmonic approach, a softening in the eigenvalues of the harmonic dynamical matrix 
\begin{equation}
\bD^{\ssz}(\bq)=\bM^{-\frac{1}{2}}\left.\frac{\partial^2 V}{\partial\bR\partial\bR}\right|_{\bRcal^{\ssz}_{\eq}}\bM^{-\frac{1}{2}} 
\label{eq:main_DH}
\end{equation}
(typically evaluated in a high-symmetry configuration) signals structural instability towards configurations with lower symmetry,
for displacements defined by the corresponding eigenvectors. 
Here $\bM$ is the mass matrix and $\bq$ is a vector of the Brillouin zone 
(we are explicitly taking advantage of crystal lattice periodicity). Notice that this approach is `classical' as 
it neglects the quantum contribution from the kinetic term of the nuclei Hamiltonian, 
i.e. it treats the nuclei as classical particles.
In general, as shown in Ref.~\citenum{PhysRevB.96.014111}, 
in order to take into account quantum, thermal and anharmonic effects in a proper way, 
we have rather to consider the phonon softening in
\begin{equation}
\bD^{\ssF}(\bq)=\bM^{-\frac{1}{2}}\left.\frac{\partial^2 F}{\partial\bRcal\partial\bRcal}\right|_{\bRcal_{\eq}}\bM^{-\frac{1}{2}}\,, 
\label{eq:main_DF}
\end{equation}
which is the (temperature-dependent) free-energy-based quantum, anharmonic generalization of the harmonic dynamical matrix.
In principle, this matrix could be evaluated by finite differences, like in a frozen-phonon approach, i.e.
calculating the variation of the SSCHA free energy (from a high-symmetry configuration) for finite displacements
of the atoms along specific distortion patterns. However, this approach would require separate SSCHA calculations
for several average atomic positions having lower symmetry. As a consequence, in order to reduce the statistical error and obtain
reliable converged results, a huge number of calculations should be performed.
%
%
On the contrary, in this paper we adopt the method introduced in Ref.~\citenum{PhysRevB.96.014111}. This approach
allows to compute the full free energy Hessian matrix with a single SSCHA calculation in the high-symmetry phase, with workload comparable 
with the one of a simple SSCHA free energy calculation. In this way, we have access to a complete picture of all the possible 
second-order lattice instabilities of the system, i.e. by diagonalizing that matrix we can find the values of the 
critical parameters and the corresponding instability patterns without any extra ad-hoc assumption 
on the expected low-symmetry phases~\cite{PhysRevB.96.014111}. 
A brief summary of the theoretical method used is reported in App.~\ref{sec:Theoretical_Method}.


We perform harmonic and SSCHA calculations for several unit cell volumes. First, we
use DFT with PBE for both the energy-force calculations used by the SSCHA 
and for the linear response calculations of the harmonic approximation.
Since we are interested in the low-temperature regime at which thermal 
fluctuations are negligible, we perform the SSCHA calculations at $0~\text{K}$, i.e. with SSCHA we 
calculate the quantum anharmonic ground-state energy.
For a given average configuration $\bRcal$, the difference between the 
Born-Oppenheimer (BO) static (i.e. classical) energy $V(\bRcal)$ and the 
SSCHA quantum (anharmonic) ground-state energy $F(\bRcal)$ defines 
the quantum (anharmonic) zero-point energy (ZPE) $E_{\ssZPE}(\bRcal)=F(\bRcal)-V(\bRcal)$.
As anticipated in Sec.~\ref{sec:Introduction},  the SSCHA has already been used to estimate the critical pressure of the $Im\bar{3}m\rightarrow R3m$ transition
in Ref.~\citenum{Errea2016}, by directly evaluating the quantum anharmonic energy difference between the high and low symmetry phases. 
In principle, such a direct approach would require the calculation of $F(\bRcal)$ for several configurations, at different volumes.
However, as said, it is very challenging and computationally demanding to reduce the statistical error of SSCHA calculations performed in the low symmetry phases. 
For that reason in Ref.~\citenum{Errea2016}, even if the SSCHA quantum energy of the high-symmetry phase 
was computed for several volumes, the SSCHA quantum energy of the low-symmetry phase was effectively computed 
only for two volumes ($\Vcal=97.85 a_0^3$, $\Vcal=102.11 a_0^3$) and, for each of them, 
only for a single $R3m$ configuration, the one corresponding to the minimum of the one dimensional double-well $Im\bar{3}m\rightarrow R3m$ distortion pattern of the BO energy. 
Subsequently, by means of fits and observing the weak volume dependence of the obtained zero-point energy curves $E_{\ssZPE}(\bRcal)$, the critical pressure
was estimated.
%
On the contrary, under the hypothesis of a second order phase transition, with the new approach no SSCHA calculations in the low-symmetry phases are required, nor
it is necessary to select \emph{a priori} specific distortion patterns or make assumptions regarding the ZPE dependence from external parameters (the volume, in this case). 
Indeed, with a computational cost which is essentially comparable with the one of the sole SSCHA quantum anharmonic energy calculation in the high-symmetry phase,
we can estimate the critical pressure with much greater precision than the previously adopted direct approach.

\begin{figure}
\centering
\includegraphics[width=\columnwidth]{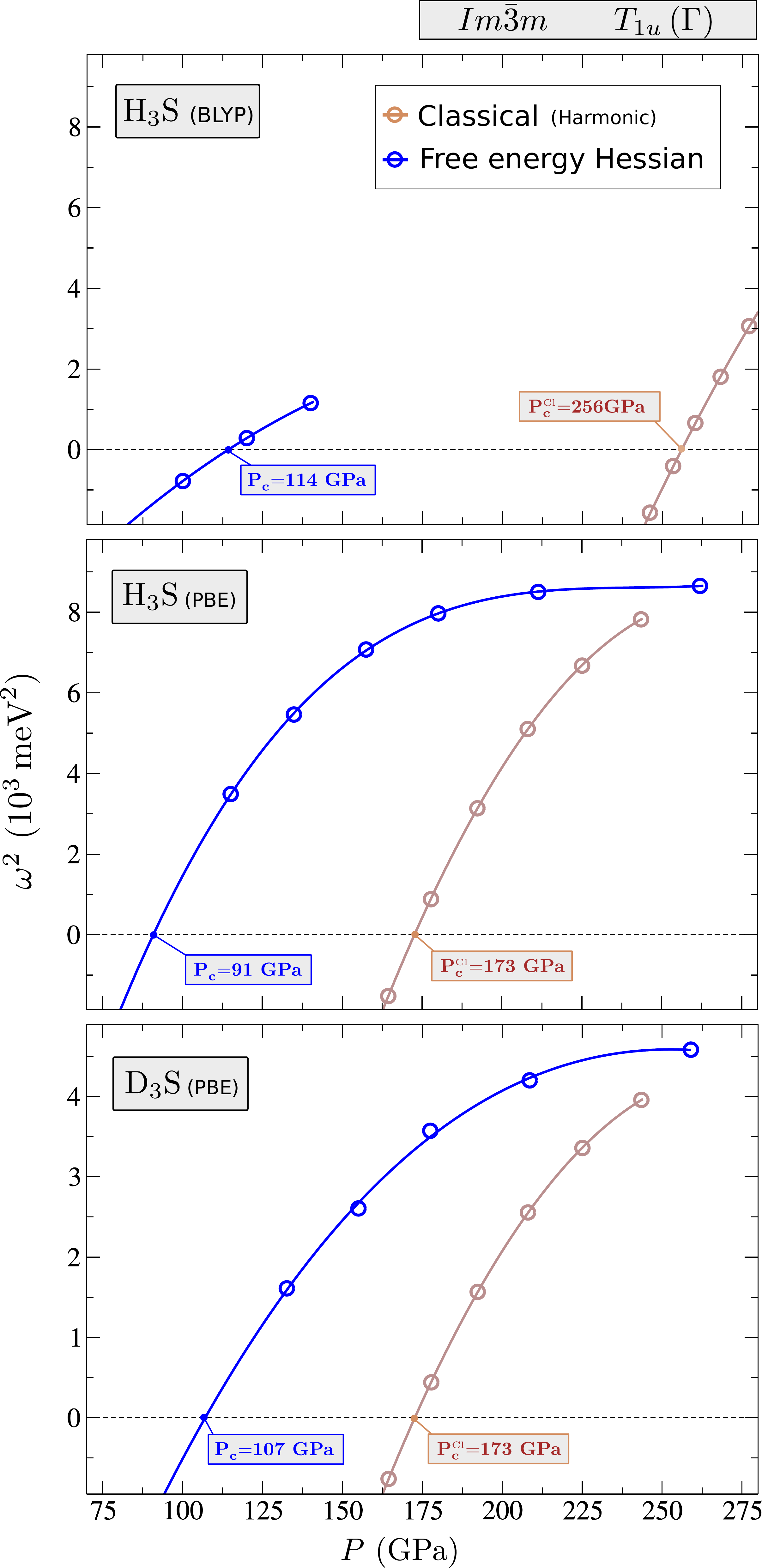}
\caption{(Color online) Squared frequency of the $T_{1u}$ mode at $\Gamma$ of $\Hsyst$ and $\Dsyst$
in the $Im\bar{3}m$ phase, as a function of the pressure. 
The `free energy Hessian' results are calculated by diagonalizing the free energy dynamical matrix $\bD^{\ssF}(\bq)$, Eq.~\eqref{eq:main_DF}. 
The `Classical' results by diagonalizing the harmonic dynamical matrix $\bD^{\ssz}(\bq)$, Eq.~\eqref{eq:main_DH}.  
The calculations are performed with PBE and BLYP for $\Hsyst$, and with PBE for $\Dsyst$, at constant volume.
Subsequently, the volume is converted to pressure by using the corresponding theoretical equation of states.
For the `free energy Hessian' results, we use the full quantum equations of states at 0 K.  
For the `Classical' result, to be consistent, we use the classical equation of states 
which does not include the zero-point energy contribution (more details in the main text). 
However, if the quantum EoS is used, the classical curve is only shifted, almost rigidly, of around +10 GPa.
For each calculation is shown the value of the critical pressure at which the mode softens
and drives the transition to the $R3m$ phase.}
\label{fig:static}
\end{figure}

At very small volume (very high pressure) both $\bRcaleq^{\ssz}$ and $\bRcaleq$, 
i.e. both classical and quantum anharmonic equilibrium configurations,  
are $Im\bar{3}m$. In other words, $\bD^{\ssz}(\bq)$ and $\bD^{\ssF}(\bq)$, 
evaluated in the high-symmetry phase $Im\bar{3}m$, are both positive-definite. 
Increasing the volume (i.e. reducing the pressure) we observe a softening in the eigenvalues of these two matrices. 
The softening is lead by an optical phonon at $\Gamma$ with irreducible representation $T_{1u}$,
which drives a second-order displacive transition to the $R3m$ phase (it can be verified through a symmetry analysis with the 
AMPLIMODES code~\cite{Orobengoa:ks5225,Perez-Mato:sh5107}, for example). 
In order to ease the comparison with experiments, in Fig.~\ref{fig:static} we show, for 
$\bD^{\ssz}(\bq)$ and $\bD^{\ssF}(\bq)$ calculated in $Im\bar{3}m$, the squared frequency 
(i.e. the eigenvalue) of the optical $T_{1u}$ mode at $\Gamma$
as a function of pressure.
We convert the unit-cell volume $\Vcal$, used for the calculations, to pressure $P$ by using the appropriate theoretical 
equation of state (EoS) at $0~\text{K}$ of Ref.~\citenum{Errea2016}. For the eigenvalues of $\bD^{\ssF}(\bq)$ we use the fully quantistic
EoS, $P(\Vcal)=-\partial F/\partial \Vcal$. For the eigenvalues of $\bD^{\ssz}(\bq)$, 
to be consistent with the fact that in this case we are neglecting quantum effects,
we use the classical EoS, $P_{\text{cl}}(\Vcal)=-\partial V/\partial \Vcal$, i.e. we discard the ZPE contribution 
(however, if we use also in this case the quantum EoS we only shift the eigenvalue curve, almost rigidly, of around +10 GPa).
The data are fitted with cubic functions in order to extrapolate the critical pressures, 
$P_c$ for $\bD^{\ssF}$ and $P_c^{\scriptscriptstyle{\text{cl}}}$ for $\bD^{\ssz}$,
at which the frequency becomes imaginary and the transition to $R3m$ is driven.

In the classical case, 
the critical pressure is identical for H${}_3$S and D${}_3$S, with
$P_c^{\scriptscriptstyle{\text{cl}}}=173$ GPa. Indeed, the harmonic dynamical matrices for the two isotopes 
differ only for the square root of the masses at denominator. Thus,
they give the transition at the same volume (and exactly at the same pressure too, since the classical equation of states without ZPE is the same for both).
When quantum effects and anharmonicity are properly 
taken into account, the results are very different, confirming the importance of going beyond the harmonic approximation. 
The critical pressures drastically reduce, with $P_c=91$ GPa and $P_c=107$ GPa for $\Hsyst$ and $\Dsyst$, respectively 
(around 10~GPa smaller than the ones presented in Ref.~\citenum{Errea2016}). 
Therefore, the $Im\bar{3}m$ phase dominates the pressure range within 
which the high $T_c$ is measured.  Moreover, a consistent isotope effect appears
with a critical pressure 16 GPa larger for D${}_3$S, which brings the result
for the heavier isotope closer to the classical limit.

%
%
In experiments the observed kink in the pressure dependence of $T_c$ on pressure release, 
which could be ascribed to the $Im\bar{3}m\rightarrow R3m$ transition,
is estimated to happen at 150 GPa and 160 GPa for $\Hsyst$ and $\Dsyst$~\cite{Drozdov2015,Einaga2016}, thus
at pressures higher than the critical pressure of the second-order $Im\bar{3}m\rightarrow R3m$ transition obtained in our calculations 
(interestingly, notice that, according to some structural research,
$\Hsyst$ at $P=$112 GPa should be already in the $C2/c$ phase~\cite{PhysRevB.93.020103}, 
thus before the $Im\bar{3}m\rightarrow R3m$ transition calculated here).
Several hypothesis can be formulated to account for this result.
%
On the calculation side, it could be tempting to ascribe the disagreement to a
failure of the self-consistent harmonic approximation (SCHA) adopted here. 
However, this is not the case, because the SCHA 
is expected to overestimate, not underestimate, the critical pressure. 
Indeed, at pressures below 173 GPa, at harmonic level the system is unstable in $Im\bar{3}m$ and stable in a $R3m$ phase, as seen. 
Therefore, at these pressures the quantum anharmonic contribution is more relevant in the $Im\bar{3}m$ 
high-symmetry phase than in the
$R3m$ low-symmetry phase. As a consequence, since the method is variational in the free energy, at these pressures 
the SCHA underestimates the free energy difference between the nonsymmetric and symmetric phases, i.e. 
underestimates the free energy curvature in the symmetric phase. 

Still on the computational side, a more reasonable hypothesis would be a failure of the DFT method 
used to perform the energy-force calculations needed by the SSCHA. 
In order to analyze this issue, we perform similar calculations for $\Hsyst$ also with the BLYP parametrization of GGA.
In this case, in order to reduce the computational cost, we perform the SSCHA
calculations on a smaller supercell ($2 \times 2 \times 2$) 
than the one ($3 \times 3 \times 3$) used for PBE. Indeed,
from the tests performed with PBE, we expect a result almost already converged
with this supercell size (details in App.~\ref{sec:Calculation_details}). 
The results, included the harmonic case, are shown in Fig.~\ref{fig:static}. 
As we can see, even if at harmonic/classical level the BLYP transition pressure increases of 83 GPa 
with respect to the PBE case, with full quantum anharmonic calculations we obtain with BLYP $P_c=114$ GPa, 
a value of the transition pressure only 23 GPa greater than the PBE one. Therefore, even with BLYP
the calculated critical pressure is very distant from the experimental measurement of the pressure at which the kink occurs. 
It is remarkable the different variation in the the critical pressure, at harmonic and anharmonic level, caused by the change of
functional used. That happens because the whole Born-Oppenheimer potential shows a strong, complex, functional dependence, 
in both the harmonic and anharmonic regimes (i.e. for both small and large displacements around the high-symmetry configuration).
This confirms, once more, the importance of taking into account anharmonic effects in order to identify, for the chosen functional,
the critical pressure at which it forms a bound quantum state that breaks the cubic symmetry.

%
%
%

On the experimental side, one possible interpretation of our results is
that the observed kink in the $T_c(P)$ curve is not related to
a $Im\bar{3}m$-$R3m$ phase transition. On lowering pressure the system in $Im\bar{3}m$ could undergo 
a very different, maybe even not simply displacive, phase transition, and different sulfur hydride stoichiometries could be involved. 
Indeed, it is worthwhile to stress that a change of the position of the H atoms is not known because the H atoms are not observable 
through x-ray scattering experiments (whereas the S atoms, as expected, appear essentially at rest through the transition, 
as previously explained).
%
%
However, another, more intriguing, hypothesis is that non-hydrostatic experimental conditions
have induced the rhombohedral transition at higher pressures. In that case, preserving the hydrostaticity during the pressure unloading, 
it could be possible to obtain the transition and, as a consequence, maintain the $T_c$ at 200 K 
even at lower pressures. 
Future experiments are thus required to shed light on the high-$T_c$ sulfur hydride behavior.

\section{Spectrum and dispersion of anharmonic phonons in the high-pressure bcc phase}
\label{sec:Spectrum_and_dispersion_of_anharmonic_phonons_in_the_high-pressure_bcc_phase}
New spectroscopic measurements in the proximity of the
observed phase transition could shed light on the structure adopted by the high-pressure superconducting sulfur hydride and 
solve the doubts discussed above.
Motivated by this consideration, in this section we use the \emph{dynamical ansatz} introduced in Ref.~\citenum{PhysRevB.96.014111} 
to calculate, with SSCHA, the single-particle spectrum of the anharmonic phonons, 
which can be probed with inelastic scattering experiments (details about the derivations in App.~\ref{sec:Theoretical_Method}).
For the energy-force calculations used by the SSCHA we utilize the PBE functional.
Here we consider $\sigma(\bq,\omega)$, the spectral function, in the point $\bq$ of the Brillouin zone, 
weighted with the factor $\omega/2\pi$. In that way, the $\omega$-integral over the real axis is normalized 
to the number of modes, and, in the pure non-interacting SSCHA case,
the spectral function is equal to a set of Dirac delta peaks
centered at $\pm\omega_\mu(\bq)$, where $\omega^2_\mu(\bq)$ are the eigenvalues of $\bD^{\ssS}(\bq)$.
Namely,
\begin{equation}
\sigma(\bq,\omega)=-\frac{\omega}{\pi}\Im\Tr{\omega^2\mathds{1}-\bD^{\ssS}(\bq)-\bPi(\bq,\omega+i0^+)}^{-1}\,,
\label{eq:maintxt_sigma}
\end{equation}
where $\bPi(\bq,z)$ is the the self-energy of the noninteracting SSCHA quasiparticles, whose dynamics is described by $\bD^{\ssS}(\bq)$.
The presence of peaks in the spectral function as a function of $\omega$, 
signals the presence of collective vibrational excitations, i.e.
phonon quasiparticles, having certain frequencies (energies). 
The sharper the peaks the more lasting are the quasiparticles, 
their life-time being inversely proportional to the peaks' width. 
Conversely, a broad spectral function 
means that anharmonicity has removed the existence of sharply defined and long-lived 
quasiparticles.
%
%

At this stage we are not making any extra assumption beyond the SSCHA itself and the dynamical \emph{ansatz}. 
If we can neglect the mixing between different SSCHA modes, we can simplify Eq.~\eqref{eq:maintxt_sigma} 
and obtain
\begin{align}
\sigma(\bq,\omega)=\sum_{\mu}\frac{1}{2}
&\left[\frac{1}{\pi}\frac{-\Im\Zcal_{\mu}(\bq,\omega)}{\left[\omega-\Re\Zcal_{\mu}(\bq,\omega)\right]^2
 +\left[\Im\Zcal_{\mu}(\bq,\omega)\right]^2}\right.\nonumber\\
+&
\left.\frac{1}{\pi}\frac{\Im\Zcal_{\mu}(\bq,\omega)}{\left[\omega+\Re\Zcal_{\mu}(\bq,\omega)\right]^2
 +\left[\Im\Zcal_{\mu}(\bq,\omega)\right]^2}\right]\,, 
\label{eq:maintxt_sigma_negelct_mixing}
\end{align}
with
\begin{equation}
\Zcal_{\mu}(\bq,\omega)=\sqrt{\omega_{\mu}^2(\bq)+\Pi_{\mu\mu}(\bq,\omega+i0^+)}\,,
\label{eq:maintxt_Zdef}
\end{equation}
where we are considering the principal value of the square root and $\Pi_{\mu\mu}(\bq,z)$ 
is the diagonal part of the self-energy for the SSCHA mode $\mu$. Therefore, mode mixing is neglected
the SSCHA modes keep their individual character in the interaction, each of them being related to a specific,
different, contribution to the total spectral function.

The form of Eq.~\eqref{eq:maintxt_sigma_negelct_mixing} resembles the
superposition of Lorentzians, but with frequency-dependent widths and shifts, 
meaning that the actual form can be quite different from the superposition of true Lorentzian functions.
However, Eq.~\eqref{eq:maintxt_sigma_negelct_mixing} in some cases can be usually expressed 
with good approximation as a sum of Lorentzians
\begin{align}
\sigma(\bq,\omega)=\sum_{\mu}\frac{1}{2}
&\left[\frac{1}{\pi}\frac{\Gamma_{\mu}(\bq)}{\left[\omega-(\omega_\mu(\bq)
  +\Delta_{\mu}(\bq))\right]^2+\left[\Gamma_{\mu}(\bq)\right]^2}\right.\nonumber\\
+&
\left.\frac{1}{\pi}\frac{\Gamma_{\mu}(\bq)}{\left[\omega+(\omega_\mu(\bq)
  +\Delta_{\mu}(\bq))\right]^2+\left[\Gamma_{\mu}(\bq)\right]^2}\right]\,,
\label{eq:maintxt_sigma_negelct_mixing_lor}
\end{align}
where 
\begin{subequations}\begin{align}
&\Delta_{\mu}(\bq)=\Re\Zcal_{\mu}(\bq,\widetilde{\omega}_{\mu}(\bq))-\omega_{\mu}(\bq)\vphantom{\frac{a}{b}}\\
&\Gamma_{\mu}(\bq)=-\Im\Zcal_{\mu}(\bq,\widetilde{\omega}_{\mu}(\bq))\vphantom{\frac{a}{b}}
\end{align}\label{eq:maintxt_shift_width_lor}\end{subequations}
are the shift (with respect to the corresponding SCHA frequency) and the half width at half maximum (HWHM) of the 
$\mu$-mode, respectively, and $\widetilde{\omega}_\mu(\bq)$ satisfies the self-consistent relation
\begin{equation}
\Re\Zcal_\mu(\bq,\widetilde{\omega}_\mu(\bq))=\widetilde{\omega}_\mu(\bq)\,.
\label{eq:maintxt_self_Z}
\end{equation}
It is worthwhile to stress that replacing Eq.~\eqref{eq:maintxt_sigma_negelct_mixing} 
with Eq.~\eqref{eq:maintxt_sigma_negelct_mixing_lor} 
is a further approximation beyond the mode-mixing negligibility. It means, by construction, that each mode  
identifies an anharmonic phonon with definite energy 
(shifted with respect to the corresponding SCHA quasi-particle energy) and lifetime.

Finally, when the SCHA self-energy is a small perturbation of the SCHA free propagator
(which, however, does not mean that we are in a perturbative regime with respect to the harmonic 
approximation), $\sigma(\bq,\omega)$ has the no mode-mixing Lorentzians form of 
Eq.~\eqref{eq:maintxt_sigma_negelct_mixing_lor} with, in particular,
\begin{subequations}\begin{align}
&\Delta_{\mu}(\bq)=\frac{1}{2\omega_{\mu}(\bq)}\Re\Pi_{\mu\mu}(\bq,\omega_{\mu}(\bq)+i0^+)\\
&\Gamma_{\mu}(\bq)=-\frac{1}{2\omega_{\mu}(\bq)}\Im\Pi_{\mu\mu}(\bq,\omega_{\mu}(\bq)+i0^+)\,.
\end{align}\label{eq:maintxt_shift_width_lor_pert}\end{subequations}

\begin{figure*}[t!]
\centering
\includegraphics[width=\textwidth]{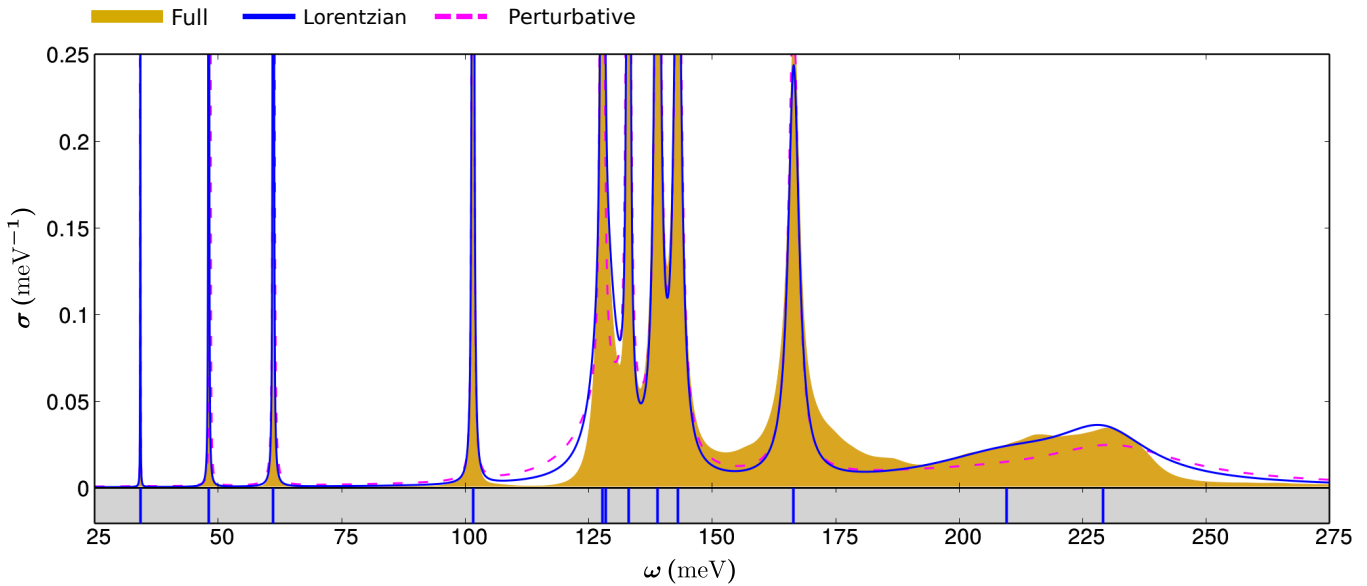}
\caption{(Color online) Phonon spectral function $\sigma(\bq,\omega)$ for $\Hsyst$ at $130$ GPa and $0$ K at $N$ 
($\bq=[0.0,0.5,-0.5]$ in $2\pi/a$ units, where $a$ is the lattice parameter). The yellow filled area indicates the result obtained with
the full formula of Eq.~\eqref{eq:maintxt_sigma}. Since the mode-mixing is negligible (for symmetry) 
the same result is obtained with Eq.~\eqref{eq:maintxt_sigma_negelct_mixing}. The blue solid line
indicates the spectrum calculated in the Lorentzian approximation  (Eqs.~\eqref{eq:maintxt_self_Z},~\eqref{eq:maintxt_sigma_negelct_mixing_lor}, 
and~\eqref{eq:maintxt_shift_width_lor}). The center of these Lorentzians define the anharmonic phonon frequencies. They are indicated with vertical lines 
in the lower panel with gray background. The dashed magenta line indicates the spectrum in the SSCHA perturbative limit, 
Eqs.~\eqref{eq:maintxt_sigma_negelct_mixing_lor} and~\eqref{eq:maintxt_shift_width_lor_pert}.}
\label{fig:dispe}
\end{figure*}

We give an example of these different level of approximations for the spectral function 
in Fig.~\ref{fig:dispe}, where we show  $\sigma(\bq,\omega)$ of $\Hsyst$ at 130 GPa, 0 K temperature, 
in the point N of the Brillouin zone, i.e. in the point $\bq=[0.0,0.5,-0.5]$ in $2\pi/a$ units, where $a$ is the lattice parameter. 
Since the spectral function is an even function of $\omega$, we plot this function only on the positive axis.
In the examined case, Eqs.~\eqref{eq:maintxt_sigma} and~\eqref{eq:maintxt_sigma_negelct_mixing} 
give the same result (yellow area), i.e. the mode-mixing can be safely neglected, because of symmetry reasons. 
However, as we can see comparing the yellow area with the solid blue line obtained with 
Eqs.~\eqref{eq:maintxt_self_Z},~\eqref{eq:maintxt_sigma_negelct_mixing_lor}, and~\eqref{eq:maintxt_shift_width_lor},
the spectrum is not Lorentzian for all the modes. Only at low energies we have very narrow peaks, well fitted by Lorentzians,
whereas at high energies the spectrum is very broad and we do not have well defined quasiparticles. 
We also show, with a dashed magenta line, the spectrum corresponding to the SSCHA perturbative approach, 
i.e. Lorentzians with shifts and widths given by 
Eqs.~\eqref{eq:maintxt_shift_width_lor_pert}. As expected, the result is not good. 
Even at low energies, where the spectrum has the form of Lorentzians superposition, 
a better result is obtained through the nonperturbative calculation of shifts and widths. 

The Lorentzian shape at low energies but a complex 
extremely broad feature at high energies are features present in the 
phonon spectrum of the high pressure $Im\bar{3}m$ phases of $\Hsyst$ and $\Dsyst$ 
throughout the Brillouin zone.
In Fig.~\eqref{fig:disp} we show the anharmonic phonon dispersion for 
$\Hsyst$ and $\Dsyst$ at three pressures in the $Im\bar{3}m$  phase and 0 K, 
along a high-symmetry path of the Brillouin zone. They are calculated through the Lorentzian
approximation, i.e. the blue lines and the blue areas indicate 
the centers $\omega_{\mu}(\bq)+\Delta_{\mu}(\bq)$ and the full widths at half maximum (FWHM) $2\Gamma_{\mu}(\bq)$ of the Lorentzians,
respectively.
The figure confirms the general trend. At low energy we have well defined phonons, i.e.
very small widths. At high energies the widths are very large and we do not have well defined quasiparticles,
the widths increasing as the critical pressure is approached. Thus, the corresponding phonons have very short lifetime and 
{decay} into the lower energy ones. These short-lived phonons are related to  
hydrogen, mainly to the modes parallel to the H--S bonds (bond stretching modes) 
as it can be seen from the projected phonon density of states in Ref.~\citenum{PhysRevLett.114.157004}.
It is also interesting to compare the anharmonic contribution to the phonon linewidth presented here with the
contribution due to the electron-phonon interaction calculated in Ref.~\citenum{PhysRevLett.114.157004}. 
Remarkably, the hydrogen bond-stretching modes also show the largest electron-phonon linewidth.
{The anharmonic contribution to the phonon linewidths of the high energy modes is similar in magnitude to the
electron-phonon coupling one (see Fig.~2 bottom panel in Ref.~\citenum{PhysRevLett.111.177002}), underlining once more the crucial role
of anharmonicity in this superconductor). }
In Fig.~\eqref{fig:disp} we also plot the $\Hsyst$ harmonic dispersion in the same phase ($Im\bar{3}m$ symmetry). 
{At the two lower pressures considered, we observe imaginary {harmonic phonons}, i.e. 
the $Im\bar{3}m$ phase is not stable in the harmonic approximation, as already seen in 
Sec.~\ref{sec:Critical_pressure_of_the_trigonal_phase_transition}, Fig.~\ref{fig:static}. 
Indeed, at these pressures the stable harmonic structure is $R3m$.

\begin{figure*}[t!]
\centering
\includegraphics[width=\textwidth]{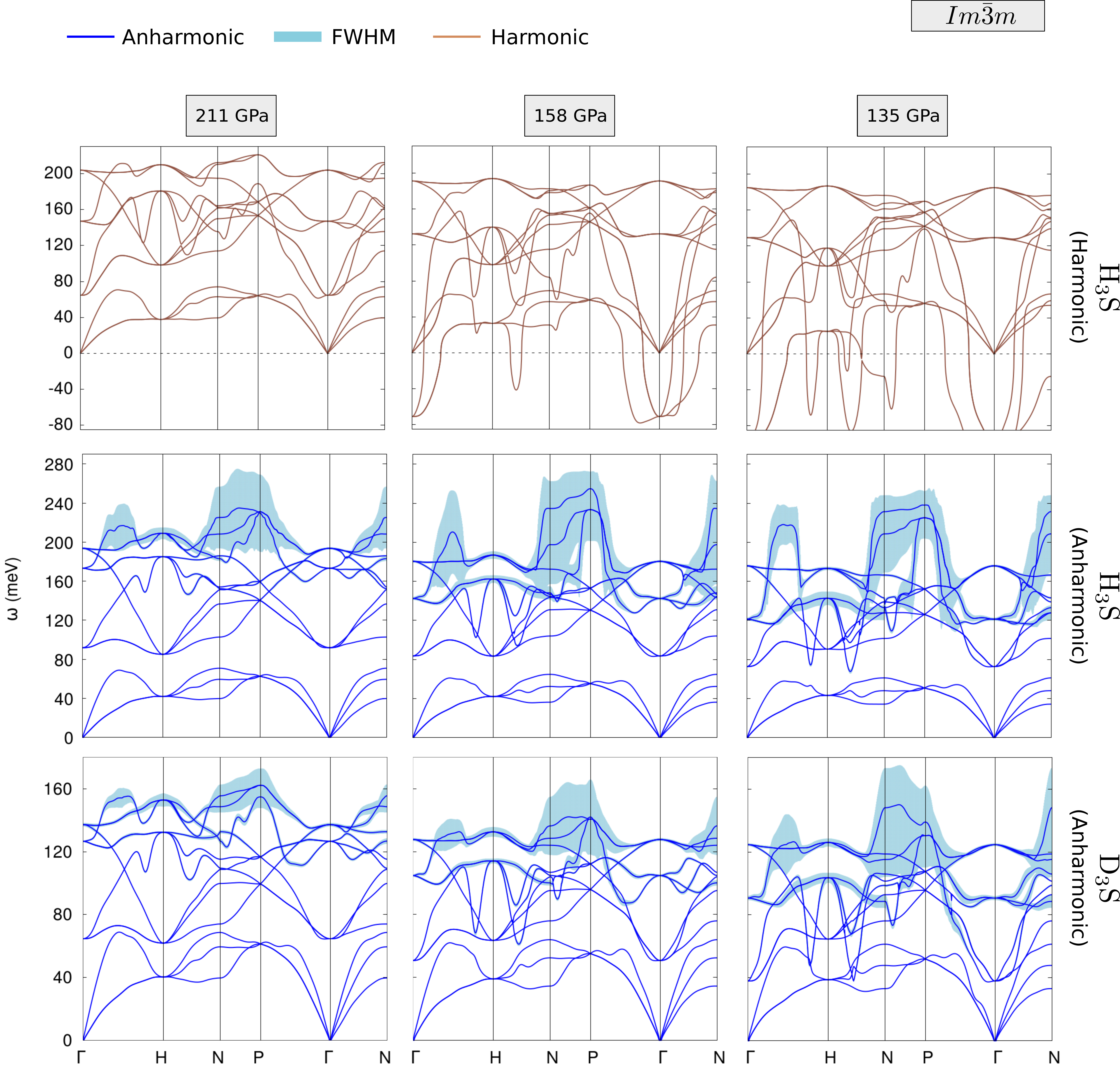}
\caption{(Color online) Harmonic and anharmonic (at 0 K) phonon dispersions for $\Hsyst$ and $\Dsyst$ 
at three pressures (same pressure
on each column) in the $Im\bar{3}m$ phase. The blue lines and the blue areas in the anharmonic plots indicate 
the centers $\omega_{\mu}(\bq)+\Delta_{\mu}(\bq)$ and the full widths at full maximum (FWHM) $2\Gamma_{\mu}(\bq)$ 
of the Lorentzians
calculated with Eqs.~\eqref{eq:maintxt_sigma_negelct_mixing_lor},~\eqref{eq:maintxt_shift_width_lor}, and~\eqref{eq:maintxt_self_Z}.} 
%
%
\label{fig:disp}
\end{figure*}

In Tables~\ref{table-gamma},~\ref{table-H}, and~\ref{table-N} we list the  
anharmonic phonon frequencies and the FWHM widths 
for the  $\Gamma=[0.0,0.0,0.0]$, H=$[-1.0,0.0,0.0]$, and N=$[0.0,0.5,-0.5]$ points 
of the Brillouin zone ($2\pi/a$ units). These results are obtained with the Lorentzian approximation, thus
the listed frequencies correspond to 
$\omega_{\mu}(\bq)+\Delta_{\mu}(\bq)$ and the full widths at half maximum (FWHM) to $2\Gamma_{\mu}(\bq)$.
Particularly interesting are the phonon energies and widths calculated at $\Gamma$, 
these being directly accessible with Raman and infrared scattering experiment~\cite{Capitani2017} 
(in the $Im\bar{3}m$ phase there are 
only infrared-active modes, in the $R3m$ phase there are Raman-active modes too). 
As we can see, in $\Gamma$ the lifetime of the anharmonic phonons remains long, even
quite close to the phase transition (linewidth at most around 2~meV), i.e. the zone center anharmonic phonons
are well-defined quasiparticles.

\begin{table*}[t!]
\caption{Anharmonic phonon frequency $\omega$ and full width at half maximum (FWHM) at $\Gamma$, 
for $\Hsyst$ and $\Dsyst$, at 0 K, at three pressures in the $Im\bar{3}m$ phase.
Infrared activity (I) for the modes is indicated. Frequencies and linewidths are given in meV units.}
\begin{center}
\begin{ruledtabular}
\begin{tabular}{cccc @{\hspace{1cm}} cc@{\hspace{1cm}}cc@{\hspace{1cm}}cc}
\multirow{2}{*}{$\Gamma=[0.0,0.0,0.0]$} &      &            &          & \multicolumn{2}{c}{\hspace{-1.0cm}135GPa} & \multicolumn{2}{c}{\hspace{-1.0cm}158GPa} & \multicolumn{2}{c}{211GPa}\\    
                       & mode & degeneracy & activity & $\omega$ & FWHM & $\omega$ & FWHM & $\omega$ & FWHM\\
\hline\\
\multirow{ 4}{*}{H${}_3$S}           
	& $\text{T}_{1u}$	&3  & I & 73.0	&	   0.2	&	  83.7	&	   0.2	&	  91.9	&	   0.2	 \\
	& $\text{T}_{1u}$	&3  & I	& 122.3	&	   1.7	&	 143.4	&	   2.3	&	 174.2	&	   1.5	 \\ 
	& $\text{T}_{2u}$	&3  & - & 175.9	&	   2.2	&	 180.5	&	   2.2	&	 193.7	&	   2.1	 \\
	&	                &   & 	&	&	  	&		&	  	&		&	         \\
 \multirow{ 3}{*}{D${}_3$S}         
	& $\text{T}_{1u}$	&3  & I	& 38.4	&	   0.1	&	  50.8	&	   0.1	&	  64.7	&	   0.0	 \\  
	& $\text{T}_{1u}$       &3  & I	& 90.9	&	   1.8	&	 105.0	&	   1.1	&	 127.0	&	   0.9	 \\ 
	& $\text{T}_{2u}$       &3  & - & 124.6	&	   1.5	&	 127.7	&	   1.5	&	 137.1	&	   2.0	 \\[1ex]
\end{tabular}
\end{ruledtabular}
\end{center}
\label{table-gamma}
\end{table*}

\begin{table*}[t!]
\caption{Anharmonic phonon frequency $\omega$ and full width at half maximum (FWHM) at $\text{H}=[-1.0,0.0,0.0]$ ($2\pi/a$ units, $a$ lattice parameter), 
for $\Hsyst$ and $\Dsyst$, at 0 K, at three pressures in the $Im\bar{3}m$ phase. Frequencies and linewidths are given in meV units.}
\begin{center}
\begin{ruledtabular}
\begin{tabular}{cc @{\hspace{1cm}} cc@{\hspace{1cm}}cc@{\hspace{1cm}}cc}
\multirow{2}{*}{$\text{H}=[-1.0,0.0,0.0]$} 
                      &             & \multicolumn{2}{c}{\hspace{-1.0cm}135GPa} & \multicolumn{2}{c}{\hspace{-1.0cm}158GPa} & \multicolumn{2}{c}{211GPa}\\    
                      & degeneracy  & $\omega$ & FWHM & $\omega$ & FWHM & $\omega$ & FWHM\\
\hline\\
\multirow{ 5}{*}{H${}_3$S}           
	& 3  & 43.2	&	   0.0	&	  42.1	&	   0.0	&	  41.9	&	   0.0	 \\   
	& 3  & 90.6	&	   0.3	&	  83.5	&	   0.1	&	  85.1	&	   0.2	 \\
	& 3  & 143.1	&	   9.7	&	 163.7	&	   6.4	&	 185.3	&	   1.4	 \\ 
	& 3  & 173.2	&	   3.1	&	 186.4	&	   5.8	&	 210.3	&	   8.8	 \\
	&    &          &	  	&		&	  	&		&	         \\
 \multirow{ 4}{*}{D${}_3$S}         
	& 3  & 38.8	&	   0.0	&	  39.1	&	   0.0	&	  40.4	&	   0.0	 \\  
	& 3  & 64.6	&	   0.0	&	  63.6	&	   0.2	&	  61.9	&	   0.0	 \\ 
        & 3  & 104.3	&	   5.9	&	 114.6	&	   2.1	&	 132.4	&	   0.9	 \\	
	& 3  & 125.9	&	   4.7	&	 132.6	&	   4.6	&	 153.6	&	   3.7	 \\[1ex]
\end{tabular}
\end{ruledtabular}
\end{center}
\label{table-H}
\end{table*}

\begin{table*}[t!]
\caption{Anharmonic phonon frequency $\omega$ and full width at half maximum (FWHM) at $\text{N}=[0.0,0.5,-0.5]$ ($2\pi/a$ units, $a$ lattice parameter), 
for $\Hsyst$ and $\Dsyst$, at 0 K, at three pressures in the $Im\bar{3}m$ phase. Frequencies and linewidths are given in meV units.}
\begin{center}
\begin{ruledtabular}
\begin{tabular}{c @{\hspace{1cm}} cc@{\hspace{1cm}}cc@{\hspace{1cm}}cc}
\multirow{2}{*}{$\text{N}=[0.0,0.5,-0.5]$} 
                        & \multicolumn{2}{c}{\hspace{-1.0cm}135GPa} & \multicolumn{2}{c}{\hspace{-1.0cm}158GPa} & \multicolumn{2}{c}{211GPa}\\    
                        & $\omega$ & FWHM & $\omega$ & FWHM & $\omega$ & FWHM\\
\hline\\
\multirow{ 12}{*}{H${}_3$S}           
&	  34.2	&	   0.0	&	  36.2	&	   0.0	&	  39.8	&	   0.0	 \\ 
&	  48.1	&	   0.0	&	  52.4	&	   0.0	&	  59.5	&	   0.0	 \\ 
&	  61.2	&	   0.0	&	  64.8	&	   0.0	&	  70.9	&	   0.1	 \\ 
&	 101.6	&	   0.2	&	 104.0	&	   0.3	&	 102.8	&	   0.4	 \\ 
&	 127.8	&	   0.3	&	 131.0	&	   0.2	&	 136.7	&	   0.4	 \\ 
&	 128.4	&	   5.1	&	 143.5	&	   1.1	& 	 151.4	&	   1.0	 \\ 
&	 133.1	&	   0.4	&	 144.7	&	   0.5	&	 152.4	&	   1.3	 \\ 
&	 138.9	&	   1.3	&	 146.6	&	   1.7	&	 154.1	&	   0.7	 \\ 
&	 143.0	&	   1.7	&	 149.5	&	   6.0	&	 186.1	&	   1.5	 \\
&	 166.5	&	   2.7	&	 172.6	&	   1.9	&	 186.2	&	   2.2	 \\ 
&	 209.6	&	  38.3	&	 186.5	&	  18.9	&	 213.2	&	  13.9	 \\ 
&	 229.1	&	  23.0	&	 240.2	&	  31.5	&	 226.2	&	  21.0	 \\ 
&	 	&	  	&	 	&	  	&	 	&	  	 \\ 
\multirow{ 11}{*}{D${}_3$S}
&	  33.1	&	   0.0	&	  34.5	&	   0.0	&	  39.8	&	   0.0	 \\
&	  48.0	&	   0.0	&	  52.5	&	   0.1	&	  59.6	&	   0.0	 \\
&	  61.4	&	   0.1	&	  64.0	&	   0.1	&	  68.7	&	   0.0	 \\
&	  76.2	&	   0.1	&	  76.0	&	   0.1	&	  74.1	&	   0.0	 \\
&	  88.6	&	   3.7	&	  95.2	&	   0.1	&	  99.8	&	   0.1	 \\
&	  92.9	&	   0.2	&	 101.0	&	   2.1	& 	 108.7	&	   0.6	 \\
&	  98.8	&	   0.8	&	 104.0	&	   0.5	& 	 109.4	&	   0.8	 \\
&	 101.5	&	   0.4	&	 104.8	&	   0.6	& 	 115.2	&	   0.5	 \\
&	 106.0	&	   0.4	&	 106.8	&	   0.3	& 	 127.7	&	   1.8	 \\ 
&	 115.6	&	  16.0	&	 124.1	&	   1.6	&	 132.7	&	   2.1	 \\
&	 118.8	&	   1.5	&	 131.4	&	  13.9	&	 151.6	&	   3.6	 \\
&	 155.1	&	  34.6	&	 139.7	&	  15.2	&	 157.5	&	   5.3	 \\[1ex]
\end{tabular}
\end{ruledtabular}
\end{center}
\label{table-N}
\end{table*}

\section{Conclusions}
\label{sec:Conclusions}

We have presented a DFT-based first-principle study of the 
high-pressure superconducting hydrogen (deuterium) sulfide, which has a record $T_c$ of 203 K at 
approximately 150 GPa, aiming to shed line on the existing experimental evidence
and clarify the phase diagram. This system, according to the majority of studies, has stoichiometry $\Hsyst$ ($\Dsyst$) and on lowering pressure,
around 150 GPa (160 GPa), undergoes a possible phase transition, as deduced from a kink in the $T_c$ evolutions as function of pressure.
According to DFT calculations, this phase transition is a rhombohedral displacement from the $Im\bar{3}m$ to the $R3m$
structure, with hydrogen bond desymmetrization and occurrence of trigonal lattice distortion of sulfur atoms.

Early XRD experiments, which can access only to the S atoms positions, observed a very small trigonal distortion,
compatibly with previous DFT calculations. However, a more recent work~\cite{PhysRevB.95.140101} that directly synthesized $\Hsyst$ 
at high pressure suggests that the $Im\bar{3}m$ transforms to the $R3m$ below 140 GPa, with a huge trigonal lattice distortion.
Our results, which show a weak functional dependence, confirm however that an $Im\bar{3}m\rightarrow R3m$ transition
would have a very small impact on the S atoms, which essentially would remain on a bcc lattice, whereas the 
largest effect would be on the position of the H atoms. Moreover, our calculations suggest that the observed large trigonal distortion,
involving the S atoms, could probably be explained by unwanted nonhydrostatic conditions in the DAC experiment.

After determining the weak rhombohedral distortion that would be associated to
the $R3m$--$Im\bar{3}m$ transition, and assuming that it is of second-order (as widely reputed),
we calculate the pressure transition $P_c$ at which it would occur in a conventional
hydrostatic setting. We use the SSCHA with DFT, to include quantum anharmonic effects, and a new method based on 
the free energy Hessian, which guarantees higher precision with respect to 
the `finite-difference' approach previously used within SSCHA.

Using PBE for the DFT energy/force calculations needed by the SSCHA, we find that the $Im\bar{3}m$ phase dominates the 
high-$T_c$ pressure range, and we estimate the $Im\bar{3}m\rightarrow R3m$ transition pressure at 91 GPa and 107 GPa 
for $\Hsyst$ and $\Dsyst$, respectively. The consistent mass effect and the difference with respect to the PBE classical harmonic
value (173 GPa) demonstrate the importance of taking into account anharmonicity and making use of a full quantum approach.

The obtained values for $P_c$ are quite different from the pressures at which the $T_c(P)$ kinks for $\Hsyst$ and $\Dsyst$ are observed.
Calculations performed with BLYP, in place of PBE, confirm the disagreement as they give $P_c=$114 GPa for $\Hsyst$. 
Several hypothesis can be formulated to account for this result. In particular, there are three main possibilities.
From the computational side, it could be a failure of the approximated exchange-correlation functionals used within DFT to compute
the energy/forces used by the SSCHA. Future Quantum Montecarlo calculations, for example, could shed light on this doubt.
On the other side, it could be a signal that the interpretation of some experimental results should be reconsidered.
For example, the observed kink in the $T_c(P)$ curve could be not related to the $Im\bar{3}m\rightarrow R3m$ transition. Maybe it could be not a simple $\Hsyst$
displacive transition, but a more complex phase transition involving different hydrogen sulfide stoichiometries. 
Another possibility, more intriguing, is that in experiments the occurrence of some anisotropy on pressure release have induced early 
the rhombohedral transition. Since the transition is associated to an abrupt decrease of the $T_c$,
keeping under control this aspect would make possible to maintain a high critical temperature even at lower pressures.

%
%
%

These considerations surely stimulate the execution of new measurements. In particular, more spectroscopic experiments 
(i.e. Raman and infrared spectroscopy) in proximity of the phase transition could shed light on the true phases involved.
In order to facilitate the comparison with future spectroscopic experiments, we used a dynamic extension of the 
SSCHA to calculate the spectrum and dispersion of anharmonic phonons for $\Hsyst$ and $\Dsyst$ 
at high-pressure in the $Im\bar{3}m$ phase, at 0 K. 
The results, among other things, confirm the huge anharmonicity of the system and that a 
nonperturbative method like the SSCHA has to be adopted. 
In general, at high energies, we find large broadening of the phonon spectra (involving H atoms modes), thus invalidating the model 
of well-defined long lasting phonon quasiparticles characterized by a finite-width Lorentzian spectrum. 
Indeed, the very low relaxation time for these modes indicates strong phonon-phonon 
scattering and decay into lower energy phonon modes. However, the vibrational spectra at zone center (accessible e.g. by infrared spectroscopy)
have very small broadening (at most linewidth around 2~meV) and anharmonic phonon quasiparticles
are well defined. 
\section*{Acknowledgments}
I.E. acknowledges financial support from the Spanish Ministry of Economy and Competitiveness (FIS2016-76617-P).
{We acknowledge the CINECA award under the ISCRA initiative (Grant
HP10BLTB9A), IDRIS, CINES and TGCC under the EDARI project A0030901202)
for the availability of high performance computing resources and support.}
R. Bianco thanks P. Barone for illuminating discussions. 
\appendix
\section{Theoretical Method}\label{sec:Theoretical_Method}
We study the lattice dynamics of $\Hsyst$ and $\Dsyst$ in the Born-Oppenheimer(BO) approximation, thus 
we consider the quantum Hamiltonian for the atoms defined by
the BO potential energy $V(\bR)$. With $\bR$ we are denoting in component-free notation
the quantity $R^{\alpha s}(\bl)$, which is a collective coordinate
that completely specifies the atomic configuration of the crystal.
The index $\alpha$ denotes the Cartesian direction,
$s$ labels the atom within the unit cell, and $\bl$ indicates the three dimensional lattice vector. 
In what follows we will also use a single composite index $a=(\alpha,s,\bl)$ to indicate Cartesian index, atom index and lattice vector together. 
Moreover, in general, we will use bold letters to indicate also other quantities in component-free notation. 

In order to take into account quantum effects and anharmonicity at a nonperturbative level,
we use the self-consistent harmonic approximation (SCHA) in its stochastic implementation~\cite{PhysRevB.89.064302,PhysRevB.96.014111}.
This is a variational mean-field method that allows to find an approximate estimation for the positional free energy $F(\Rcal^{\alpha s}(\bl))$, 
i.e. the free energy of the crystal as a function of the average atomic position $\Rcal^{\alpha s}(\bl)$ 
(the \emph{centroids}). For a given centroid $\bRcal$, the SCHA free energy is obtained
through an auxiliary quadratic Hamiltonian, the SCHA Hamiltonian $\Hcal_{\bRcal}$, that minimizes an opportune functional.
In particular, this allows to find $F(\bRcal;\xi)$, the positional free energy as a function of an external parameter $\xi$ like, 
for example, the temperature or, as in the case examined in this paper, the volume.
Given $\xi$, the equilibrium configuration $\bRcaleq(\xi)$ corresponds to the minimum of $F(\bRcal;\xi)$ with respect to $\bRcal$.
In a displacive second-order phase transition, according to Landau's theory~\cite{landau2013statistical}, 
there is a critical value $\xi_c$ such that at $\xi > \xi_c$ the  equilibrium configuration $\bRcaleq(\xi)$ is in a high symmetry configuration $\bRcalhs$ but,
on lowering the value of the parameter, $\bRcalhs$ becomes a saddle point at $\xi_c$ and $\bRcaleq(\xi)$ moves to a lower symmetry configuration. 
As a consequence, the free energy Hessian evaluated in $\bRcalhs$, $\left.\partial^2F/\partial\bRcal\partial\bRcal\right|_{\bRcalhs;\xi}$, 
at $\xi>\xi_c$ is positive definite but it develops one or multiple negative eigendirections at $\xi_c$. 
Therefore, by evaluating the free energy Hessian in $\bRcalhs$ and studying its spectrum as a function of $\xi$, we can  predict 
the occurrence of a displacive second-order phase transition, estimate the critical value $\xi_c$, and 
calculate the distortion pattern that reduces the free energy by considering the
negative eigendirections. For notation clarity, from now on the dependence on $\xi$ will be understood. 

The SCHA free energy Hessian in a centroid $\bRcal$ can be computed by using the analytic formula (in component-free notation)~\cite{PhysRevB.96.014111}
\begin{equation}
\frac{\partial^2 F}{\partial\bRcal\partial\bRcal}
=\bPhi+\overset{\sst}{\bPhi}\bLambda(0)
\left[\mathds{1}-\overset{\ssf}{\bPhi}\bLambda(0)\right]^{-1}\overset{\sst}{\bPhi},
\label{eq:Hessian}
\end{equation}
with
\begin{gather}
\bPhi=\Bavg{\frac{\partial^2 V}{\partial \bR\partial \bR}}_{\rho_{\Hcal_{\bRcal}}},\nonumber\\
\,\nonumber\\
\overset{\sst}{\bPhi}=\avg{\frac{\partial^3 V}{\partial \bR\partial \bR\partial \bR}}_{\rho_{\Hcal_{\bRcal}}},\quad
\overset{\ssf}{\bPhi}=\avg{\frac{\partial^4 V}{\partial \bR\partial \bR\partial \bR \partial \bR}}_{\rho_{\Hcal_{\bRcal}}},
\label{eq:smeared_FC}
\end{gather}
where the averages are with respect to the density matrix of the SCHA Hamiltonian
$\Hcal_{\bRcal}$, i.e. $\rho_{\Hcal_{\bRcal}}=e^{-\beta\Hcal_{\bRcal}}/\text{tr}\left[e^{-\beta\Hcal_{\bRcal}}\right]$, 
and $\beta=(k_bT)^{-1}$ where $k_b$ is the Boltzmann constant and $T$ is the temperature. 
In Eq.~\eqref{eq:Hessian} the value at $z=0$ of the 4th-order
tensor $\bLambda(z)$ is used. For a generic complex number $z$, this tensor is defined, in components, by 
\begin{align}
&\Lambda^{abcd}(z)=-\frac{1}{2}\sum_{\mu\nu}\,F(z,\omega_\mu,\omega_\nu)\nonumber\\
&\times
\sqrt{\frac{\hbar}{2M_a\omega_{\mu}}}e_{\mu}^a\,\sqrt{\frac{\hbar}{2M_b\omega_{\nu}}}e^b_{\nu}\,
\sqrt{\frac{\hbar}{2M_c\omega_{\mu}}}e^c_{\mu}\,\sqrt{\frac{\hbar}{2M_d\omega_{\nu}}}e^d_{\nu},
\end{align}
with $M_a$ the mass of the atom $a$, $\omega^2_{\mu}$ and $e^a_{\mu}$ eigenvalues and corresponding eigenvectors of the matrix 
$\Phi_{ab}/\sqrt{M_aM_b}$, respectively, and
\begin{align}
F(z,\omega_\nu,\omega_{\mu})=&\frac{2}{\hbar}\left[\frac{(\omega_{\mu}+\omega_{\nu})[1+n_{\B}(\omega_{\mu})+n_{\B}(\omega_{\nu})]}{(\omega_\mu+\omega_\nu)^2-z^2}\right.\nonumber\\
&-\left.\frac{(\omega_{\mu}-\omega_{\nu})[n_{\B}(\omega_{\mu})-n_{\B}(\omega_{\nu})]}{(\omega_\mu-\omega_\nu)^2-z^2}\right]\,,
\end{align}
where $n_{\B}(\omega)=1/(e^{\beta\hbar\omega}-1)$ is the bosonic occupation number.
Expanding the geometric series in Eq.~\eqref{eq:Hessian} and retaining only the first term beyond
the SCHA matrix $\bPhi$, i.e. the so called \emph{bubble} $\overset{\sst}{\bPhi}\bLambda(0)\overset{\sst}{\bPhi}$, we can write
\begin{equation}
\frac{\partial^2 F}{\partial\bRcal\partial\bRcal}\simeq\bPhi+\overset{\sst}{\bPhi}\bLambda(0)\overset{\sst}{\bPhi}\,.
\label{eq:hessian_w_bubble}
\end{equation}
In the cases studied here, this is sufficient to obtain converged results (App.~\ref{sec:Calculation_details}).

Dividing the free energy Hessian in the equilibrium configuration by the square root of the masses we define 
\begin{equation}
\bD^{\ssF}=\bM^{-\frac{1}{2}}\left.\frac{\partial^2F}{\partial\bRcal\partial\bRcal}\right|_{\bRcaleq}\bM^{-\frac{1}{2}}\,,
\label{eq:app_DF}
\end{equation}
where we have indicated with $M_{ab}=\delta_{ab}M_a$ the mass matrix. 
A lattice instability corresponds to a softening of an eigenvalue of $\bD^{\ssF}$. $\bD^{\ssF}$ is
the quantum, anharmonic, temperature-dependent generalization of the standard temperature-independent harmonic dynamical matrix
\begin{equation}
\bD^{\ssz}=\bM^{-\frac{1}{2}}\left.\frac{\partial^2V}{\partial\bR\partial\bR}\right|_{\bRcaleq^{\ssz}}\bM^{-\frac{1}{2}}\,,
\end{equation}
where $\bRcaleq^{\ssz}$ is the temperature-independent minimum of $V(\bR)$. Notice that the free energy dynamical matrix $\bD^{\ssF}$
is obtained from the positive-definite (by construction) auxiliary `SSCHA dynamical matrix'
\begin{equation}
\bD^{\ssS}=\bM^{-\frac{1}{2}}\left.\bPhi\right|_{\bRcaleq}\bM^{-\frac{1}{2}}\,,
\label{eq:app_DS}
\end{equation}
plus a second term, see Eq.~\eqref{eq:Hessian}. It is this second term that allows to have zero and negative eigenvalues (i.e.instabilities). 

Since in this work we are considering a crystal, we can take advantage of lattice periodicity and Fourier transform
the relevant quantities with respect to the lattice indices. That allows to make independent analysis 
for each $\bq$ point in the Brillouin zone.
We define the Fourier transform of the atomic position as
\begin{equation}
R^{\alpha s}(\bq)=\frac{1}{N_c}\sum_{\bl}e^{-i\bq\bl}R^{\alpha s}(\bl)\,,
\end{equation}
and the Fourier transform of the SCHA force constants as
\begin{align}
&\overset{\ssn}{\Phi}_{\alpha_1s_1\ldots\alpha_ns_n}(\bq_1,\ldots,\bq_n)=\nonumber\\
&\,\frac{1}{N_c}\sum_{\bl_1\ldots\bl_n}e^{i\left(\bq_1\bl_1+\cdots+\bq_n\bl_n\right)}\overset{\ssn}{\Phi}_{\alpha_1s_1\ldots\alpha_ns_n}(\bl_1,\ldots,\bl_n)\,,
\end{align}
where with $\overset{\ssn}{\bPhi}$ we are indicating the objects defined in Eq.~\eqref{eq:smeared_FC} for a generic integer $n$ (however, notice that
for $n=2$ we did not use any superscript in that definition). $N_c$ is the number of unit-cells $\bl$ comprising the supercell considered, 
i.e. the number of $\bq$ points of the 
corresponding commensurate mesh in the first Brillouin zone of the reciprocal space.
Therefore, in the H${}_3$S case considered here, since there are $4$ atoms in the unit cell,
for each point $\bq$ the $\bD^{\ssz}(\bq)$, $\bD^{\ssS}(\bq)$ and $\bD^{\ssF}(\bq)$ are $12\times12$ matrices. 
We indicate with $\omega_{\mu}^2(\bq)$ and $e^a_{\mu}(\bq)$ the eigenvalues and corresponding eigenvectors
of $\bD^{\ssS}(\bq)$, respectively. 
Diagonalizing the free energy dynamical matrix $\bD^{\ssF}(\bq)$, we can study a structural instability with respect to a distortion having the lattice modulation characterized by the pseudomomentum $\bq$. In Fig.~\eqref{fig:SSCHA_role_static} we show the different role played by 
$\bD^{\ssF}(\bq)$ and the positive-definite $\bD^{\ssS}(\bq)$. 
We consider $\Hsyst$ and the eigenvalue of these two matrices for the $T_{1u}$ mode in $\Gamma$, 
calculated with BLYP (see Fig.~\ref{fig:static} in 
Sec.~\ref{sec:Critical_pressure_of_the_trigonal_phase_transition}). 
By construction, $\bD^{\ssS}(\bq)$ cannot have negative eigenvalues, whereas $\bD^{\ssF}(\bq)$ has 
zero eigenvalue in correspondence of a second-order structural instability.

\begin{figure}
\centering
\includegraphics[width=\columnwidth]{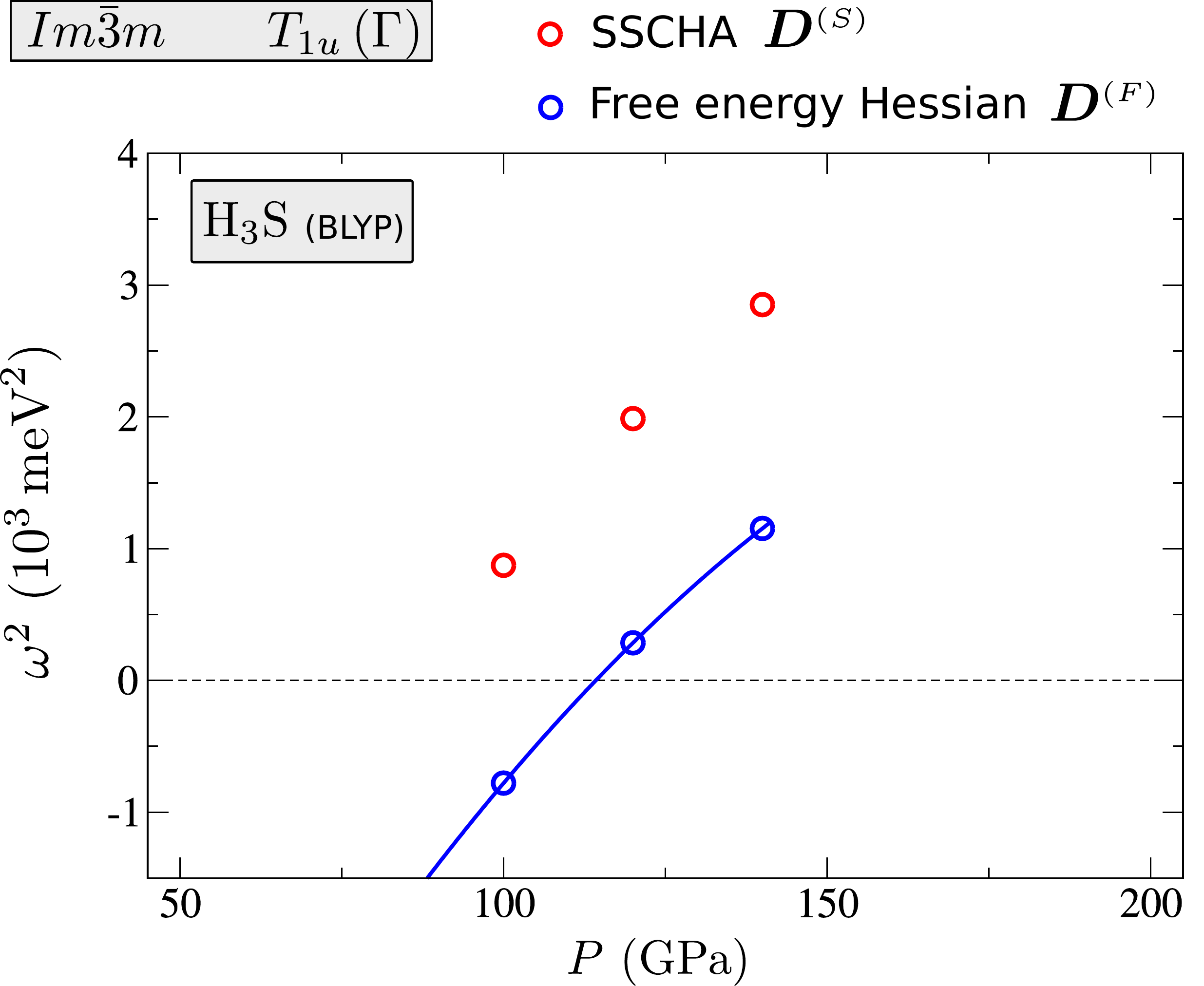}
\caption{(Color online) Squared frequency of the $T_{1u}$ mode
at $\Gamma$ of $\Hsyst$ in the $Im\bar{3}m$ phase, as a function of
the pressure, for $\bD^{\ssF}$, Eq.~\eqref{eq:app_DF}, and $\bD^{\ssS}$, Eq.~\eqref{eq:app_DS}, calculated with BLYP.
While $\bD^{\ssS}$ is positive-definite and thus cannot have negative eigenvalues, $\bD^{\ssF}$ has zero eigenvalues
at the critical pressure for which the mode drives a transition to the $R3m$ phase.}
\label{fig:SSCHA_role_static}
\end{figure}

As shown in Ref.~\citenum{PhysRevB.96.014111}, in the context of the SCHA it is possible to 
formulate an \emph{ansatz} to give an approximate expression to the one-phonon Green function $\bG(z)$ for the
variable $\sqrt{M_a}(R^a-\Rcaleq^a)$
\begin{equation}
\bG^{-1}(z)=z^2\mathds{1}-\left(\bD^{\ssS}+\bPi(z)\right)\,,
\label{Eq:Gm1}
\end{equation}
where $\bPi(z)$ is the SCHA self-energy, given by
\begin{equation}
\bPi(z)=\bM^{-\frac{1}{2}}\,\overset{\sst}{\bPhi}\bLambda(z)
\left[\mathds{1}-\overset{\ssf}{\bPhi}\bLambda(z)\right]^{-1}\overset{\sst}{\bPhi}\,\bM^{-\frac{1}{2}}.
\label{Eq:app_Pi}
\end{equation}
Correctly, in the static limit we recover the free energy dynamical matrix
(Cfr.~Eqs.~\eqref{eq:Hessian},~\eqref{eq:app_DF} with Eqs.~\eqref{Eq:Gm1},~\eqref{Eq:app_Pi}):
\begin{equation}
-\bG^{-1}(0)=\bD^{\ssS}+\bPi(0)=\bD^{\ssF}\,.
\label{eq:static_limit}
\end{equation}
Analogously to the free energy Hessian, Eq.~\eqref{Eq:app_Pi}
can be expanded in series and in the present paper we find sufficient to retain only the first term, 
the so called \emph{bubble} $\overset{\ssB}{\bPi}(z)$:
\begin{equation}
\bPi(z)\approx\overset{\ssB}{\bPi}(z)=\bM^{-\frac{1}{2}}\,\overset{\sst}{\bPhi}\bLambda(z)\overset{\sst}{\bPhi}\,\bM^{-\frac{1}{2}}.
\label{Eq:app_PiB}
\end{equation}
From $\bG(z)$ we calculate the one-phonon spectral function $-2\Im\Tr{\bG(\omega+i0^+)}$. Peaks
in the spectral function as a function of $\omega$ signal the presence of  collective vibrational excitations, i.e. phonon quasiparticles, having certain frequencies (energies).
The sharper are the peaks the more lasting are these quasiparticles, their life-time being inversely proportional to the width, 
whereas a broad spectral function means that anharmonicity  has removed the existence of quasiparticles with definite identity. 
These are the kind of information that can be probed with inelastic scattering experiments, for example.

We take advantage of the lattice periodicity also in this case and, Fourier transforming with respect to the lattice indices, 
we consider separated spectral functions $-2\Im\Tr{\bG(\bq,\omega+i0^+)}$ for each $\bq$ point in the Brillouin zone.
Moreover, we find convenient to consider the spectral function \emph{weighted} with the factor $\omega/2\pi$,
\begin{align}
\sigma(\bq,\omega)&=-\frac{\omega}{\pi}\Im\Tr{\bG(\bq,\omega+i0^+)}\\
              &=-\frac{\omega}{\pi}\Im\Tr{\omega^2\mathds{1}-\left(\bD^{\ssS}(\bq)+\bPi(\bq,\omega+i0^+)\right)}^{-1}\,,
\label{eq:sigma_2}
\end{align}
because its $\omega$-integral over the real axis gives the total number of modes and, in the zero self-energy case,
it gives equal Dirac-delta peaks at $\pm\omega_{\mu}(\bq)$, where $\omega_\mu(\bq)$ are the noninteracting SCHA boson frequencies. 

Evaluation of Eq.~\eqref{eq:sigma_2} requires the inversion of a matrix, which in some cases can be computationally demanding.
However, this is not necessary if the \emph{mode mixing} can be neglected, i.e. if the coupling between different eigenmodes
of $\bD^{\ssS}(\bq)$, induced by the self-energy $\bPi(\bq,\omega+i0^+)$, can be neglected:
\begin{equation}
\Pi_{\mu\nu}(\bq,\omega+i0^+)\simeq\delta_{\mu\nu}\Pi_{\nu\nu}(\bq,\omega+i0^+)\,,
\end{equation}
where $\Pi_{\mu\nu}(\bq,z)=\sum_{ab}\Pi_{ab}(\bq,z)e^a_{\mu}(-\bq)e^b_{\nu}(\bq)$ is the SCHA self-energy in the basis of the SCHA-modes. In that case
\begin{align}
&\Tr{\omega^2\mathds{1}-\left(\bD^{\ssS}(\bq)+\bPi(\bq,\omega+i0^+)\right)}^{-1}\nonumber\\
&\qquad\qquad=\sum_{\nu}\frac{1}{\omega^2-\omega_{\nu}^2(\bq)-\Pi_{\nu\nu}(\bq,\omega+i0^+)}\,,
\label{eq:tr_negelct_mixing}
\end{align}
and, with some manipulations, from Eq.~\eqref{eq:sigma_2} we obtain
\begin{align}
\sigma(\bq,\omega)=\sum_{\mu}\frac{1}{2}
&\left[\frac{1}{\pi}\frac{-\Im\Zcal_{\mu}(\bq,\omega)}{\left[\omega-\Re\Zcal_{\mu}(\bq,\omega)\right]^2
 +\left[\Im\Zcal_{\mu}(\bq,\omega)\right]^2}\right.\nonumber\\
+&
\left.\frac{1}{\pi}\frac{\Im\Zcal_{\mu}(\bq,\omega)}{\left[\omega+\Re\Zcal_{\mu}(\bq,\omega)\right]^2
 +\left[\Im\Zcal_{\mu}(\bq,\omega)\right]^2}\right]\,, 
\label{eq:sigma_negelct_mixing}
\end{align}
with 
\begin{equation}
\Zcal_{\mu}(\bq,\omega)=\sqrt{\omega_{\mu}^2(\bq)+\Pi_{\mu\mu}(\bq,\omega+i0^+)}\,,
\label{eq:Zdef}
\end{equation}
where we are considering the principal value of the square root.
Thus, the negligibility of the mixing-mode maintains separated the individuality of the
SCHA modes in the interaction, each of them being related to a specific, different, contribution to the total spectral function.

The form of Eq.~\eqref{eq:sigma_negelct_mixing} resembles a superposition of Lorentzians, but with frequency-dependent 
widths and shifts, meaning that 
the actual form, in general, can be quite different from the superposition of true Lorentzian functions.
However, considered a $\widetilde{\omega}_\mu(\bq)$ that satisfy the self-consistent relation
\begin{equation}
\Re\Zcal_\mu(\bq,\widetilde{\omega}_\mu(\bq))=\widetilde{\omega}_\mu(\bq)\,,
\label{eq:self_Z}
\end{equation}
Eq.~\eqref{eq:sigma_negelct_mixing} can be usually expressed with good approximation as a sum of Lorentzians
\begin{align}
\sigma(\bq,\omega)=\sum_{\mu}\frac{1}{2}
&\left[\frac{1}{\pi}\frac{\Gamma_{\mu}(\bq)}{\left[\omega-(\omega_\mu(\bq)
  +\Delta_{\mu}(\bq))\right]^2+\left[\Gamma_{\mu}(\bq)\right]^2}\right.\nonumber\\
+&
\left.\frac{1}{\pi}\frac{\Gamma_{\mu}(\bq)}{\left[\omega+(\omega_\mu(\bq)
  +\Delta_{\mu}(\bq))\right]^2+\left[\Gamma_{\mu}(\bq)\right]^2}\right]\,,
\label{eq:sigma_negelct_mixing_lor}
\end{align}
where 
\begin{subequations}\begin{align}
&\Delta_{\mu}(\bq)=\Re\Zcal_{\mu}(\bq,\widetilde{\omega}_{\mu}(\bq))-\omega_{\mu}(\bq)\vphantom{\frac{a}{b}}\\
&\Gamma_{\mu}(\bq)=-\Im\Zcal_{\mu}(\bq,\widetilde{\omega}_{\mu}(\bq))\vphantom{\frac{a}{b}}
\end{align}\label{eq:shift_width_lor}\end{subequations}
are shift (with respect to the SCHA frequency) and half width at half maximum (HWHM) of the $\mu$-mode, respectively. 
Replacing Eq.~\eqref{eq:sigma_negelct_mixing} with Eq.~\eqref{eq:sigma_negelct_mixing_lor} 
is a further approximation beyond the mode-mixing negligibility. It implies that each mode  
identifies an anharmonic phonon with definite energy (shifted with respect to the corresponding SCHA quasiparticle energy) 
and lifetime (inverse of the Lorentzian width).

Finally, we consider the case in which the SCHA self-energy is
a small perturbation of the SCHA free propagator (it is worthwhile to stress that this does not mean that we are in a perturbative regime with respect to the harmonic 
approximation). According to perturbation theory, in that case at lowest order $\omega_{\nu}^2(\bq)+\Pi_{\nu\nu}(\bq,\omega+i0^+)$
are eigenvalues of $\bD^{\ssS}(\bq)+\bPi(\bq,\omega+i0^+)$, thus 
Eqs.~\eqref{eq:tr_negelct_mixing},~\eqref{eq:sigma_negelct_mixing},~\eqref{eq:Zdef} follow with, in particular,
\begin{equation}
\Zcal_{\mu}(\bq,\omega)\simeq\omega_{\mu}(\bq)+\frac{\Pi_{\mu\mu}(\bq,\omega+i0^+)}{2\omega_{\mu}(\bq)}\,.
\label{eq:Zdef_pert}
\end{equation}
Moreover, since $\Pi_{\mu\mu}(\bq,\omega+i0^+)\ll\omega_{\mu}(\bq)$, the self-consistent relation Eq.~\eqref{eq:self_Z} is satisfied with
the SCHA frequency, i.e. $\widetilde{\omega}_{\mu}(\bq)=\omega_{\mu}(\bq)$. Therefore, in the SCHA perturbative limit we have 
Eq.~\eqref{eq:sigma_negelct_mixing_lor} with shift and HWHM  given by
\begin{subequations}\begin{align}
&\Delta_{\mu}(\bq)=\frac{1}{2\omega_{\mu}(\bq)}\Re\Pi_{\mu\mu}(\bq,\omega_{\mu}(\bq)+i0^+)\\
&\Gamma_{\mu}(\bq)=-\frac{1}{2\omega_{\mu}(\bq)}\Im\Pi_{\mu\mu}(\bq,\omega_{\mu}(\bq)+i0^+)\,,
\end{align}\label{eq:shift_width_lor_pert}\end{subequations}
respectively.

We conclude this section by showing that the SSCHA dynamical matrix is only an auxiliary quantity, which in general
cannot be used to describe dynamic properties, i.e. it does not describe the physical phonons. In Fig~\ref{fig:SSCHA_role_dynamic}, 
for $\Hsyst$ in the $Im\bar{3}m$ phase, at 0 K and 185 GPa, we show the 
the dispersion curves of the SCHA matrix $\bD^{\ssS}(\bq)$, i.e. the SCHA frequencies $\omega_{\mu}(\bq)$,
and true phonon dispersion in the Lorentzian approximation, 
$\omega_{\mu}(\bq)+\Delta_{\mu}(\bq)$, with linewidth $2\Gamma_{\mu}(\bq)$, as it is calculated with 
Eqs.~\eqref{eq:Zdef},~\eqref{eq:self_Z}, and~\eqref{eq:shift_width_lor} (see also Fig.~\ref{fig:disp}.
in Sec.~\ref{sec:Spectrum_and_dispersion_of_anharmonic_phonons_in_the_high-pressure_bcc_phase}). 
The true phonon dispersion and the SCHA dispersion are very different. 
Moreover notice that, by construction, no linewidth is associated to the SCHA dispersion curves (the SCHA quasiparticles are non interacting, 
i.e. we are discarding the self-energy, thus there is zero linewidth, i.e. they have infinite lifetime).

\begin{figure}[t!]
\centering
\includegraphics[width=\columnwidth]{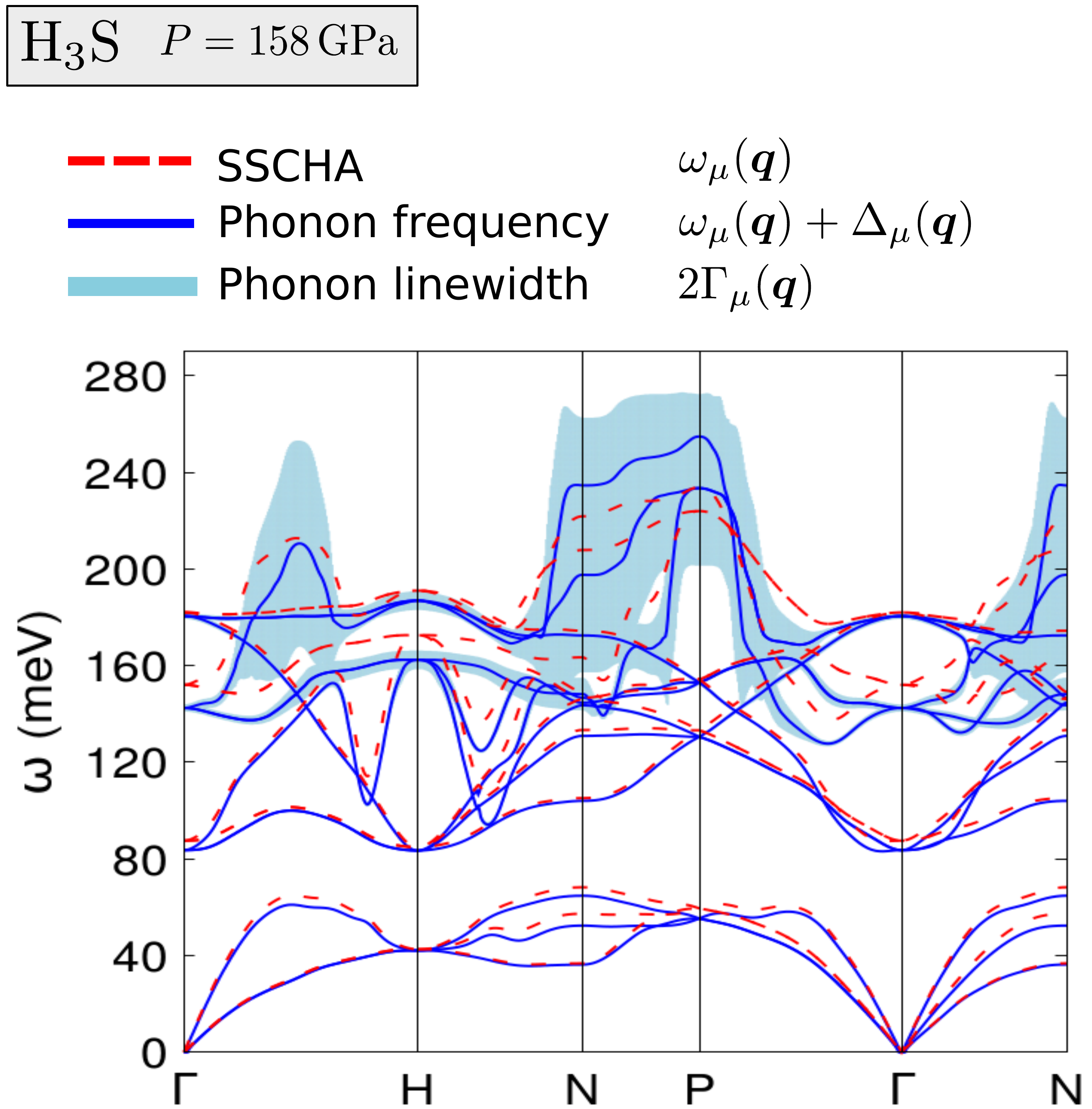}
\caption{(Color online). $\Hsyst$ in the $Im\bar{3}m$ phase, at 0 K and 185 GPa. SCHA dispersion curves, $\omega_{\mu}(\bq)$,
anharmonic phonon dispersion curves, $\omega_{\mu}(\bq)+\Delta_{\mu}(\bq)$, and linewidths, $2\Gamma_{\mu}(\bq)$,
calculated with Eqs.~\eqref{eq:tr_negelct_mixing},~\eqref{eq:sigma_negelct_mixing},~\eqref{eq:Zdef}.}
\label{fig:SSCHA_role_dynamic}
\end{figure}

\section{Calculation details}\label{sec:Calculation_details}
We performed plane-wave density-functional theory~\cite{parr1994density,martin2004electronic,gross1995density} (DFT) 
and density-functional perturbation theory~\cite{RevModPhys.73.515} (DFPT) 
calculations using the QUANTUM-ESPRESSO package~\cite{0953-8984-21-39-395502}.
We used the generalized gradient approximation (GGA) for the exchange-correlation functional~\cite{PhysRevLett.77.3865}, under the Perdew-Burke-Ernzherhof (PBE)
and the Becke-Lee-Yang-Parr~\cite{PhysRevA.38.3098,PhysRevB.37.785} (BLYP) parametrizations. We used ultrasoft pseudopotentials~\cite{PhysRevB.41.7892}, 
a plane-wave cut-off energy of 60 Ry for the kinetic energy and 600 Ry for the charge density, and a Hermite-Gaussian smearing of 0.04 Ry. 
For the unit-cell calculation, the integration in reciprocal space  was performed on a $32 \times 32 \times 32$ Monkhorst-Pack grid~\cite{PhysRevB.13.5188} 
of the Brillouin zone (BZ). This mesh was adjusted accordingly in the supercell calculations. 
The self-consistent solution of the Kohn-Sham equations was obtained when the total energy changed by
less than $10^{-9}$ Ry. The harmonic phonon dispersion was obtained with Fourier interpolation after DFPT calculation performed on a $12\times 12 \times 12$ 
Monkhorst-Pack grid of the BZ. The harmonic ZPE contribution reported in Sec.~\ref{sec:Effect_of_nonhydrostatic_pressure}
has been calculated with Fourier interpolation after DFPT calculation performed on a $6\times 6 \times 6$ 
Monkhorst-Pack grid of the BZ.

\begin{figure}[t!]
\centering
\includegraphics[width=\columnwidth]{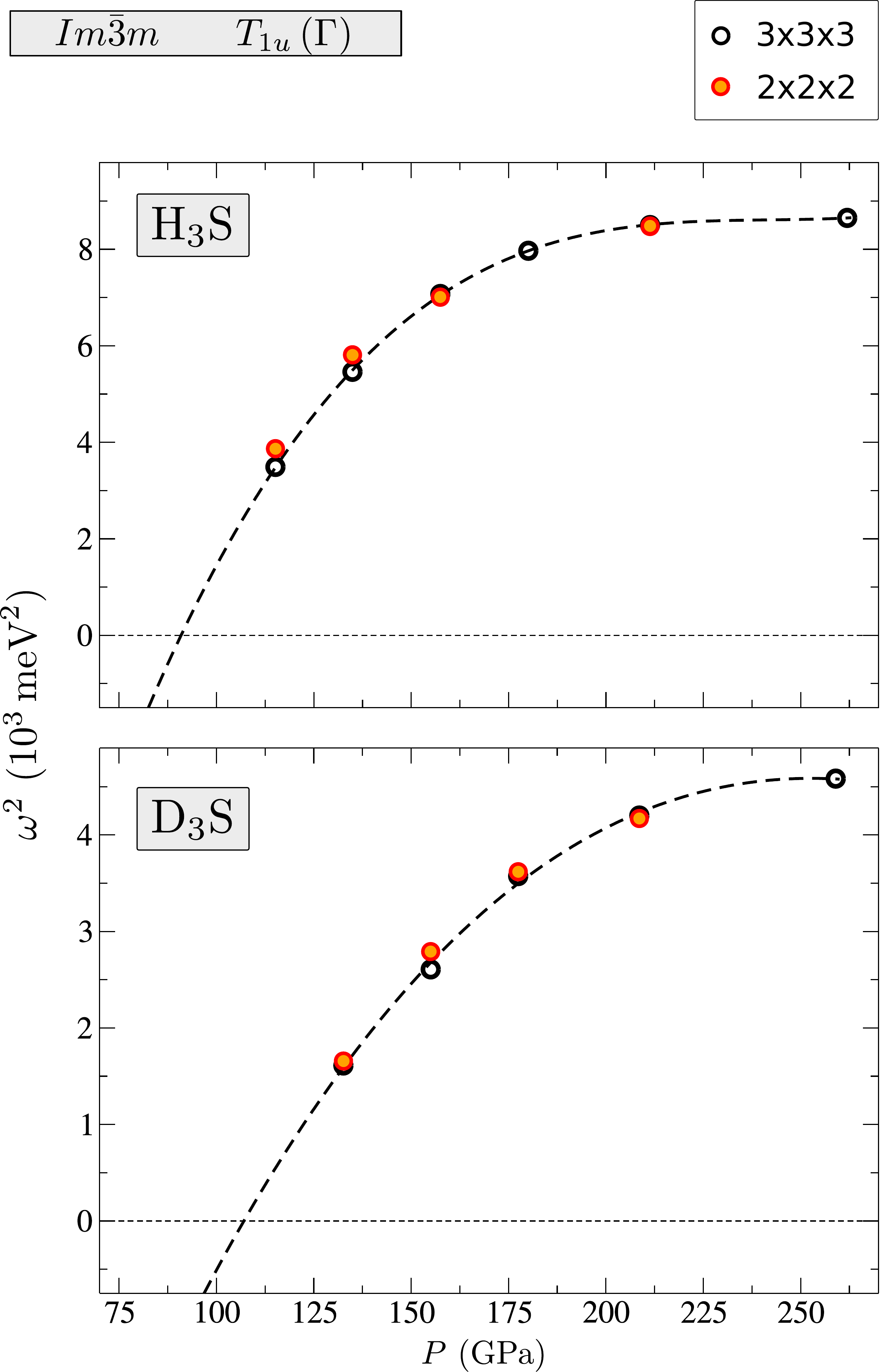}
\caption{(Color online) Squared $T_{1u}$ optical mode frequency at $\Gamma$ of the free energy dynamical matrix $\bD^{\ssF}$
as a function of pressure for $\Hsyst$ (upper panel) and $\Dsyst$ (lower panel), 
using a $2 \times 2 \times 2$ (full dots) and $3 \times 3 \times 3$ (empty dots) 
supercell in the SSCHA calculation. In both cases, 
the static bubble used to compute $\bD^{\ssF}$ with $\bD^{\ssS}$ through Eq.~\eqref{eq:static_limit}, is 
calculated with Eq.~\eqref{eq:app_Pi_interp}, integrating on a $12 \times 12 \times 12$ grid.}
\label{fig:static_conv}
\end{figure}

The stochastic self-consistent harmonic approximation (SSCHA) calculations~\cite{PhysRevB.89.064302,PhysRevB.96.014111} were performed 
using $3\times 3 \times 3$ (for PBE) and $2\times 2 \times 2$ (for BLYP) supercells,
and populations of the order of $10^3$ elements. For the self-energy $\bPi(z)$, Eq.~\eqref{Eq:app_Pi}, we considered only the first term in the expansion of the geometric series, i.e.
we considered only the \emph{bubble} $\overset{\ssB}{\bPi}(z)$, Eq.~\eqref{Eq:app_PiB}. 
Indeed, we verified that, in our case,
the terms beyond the bubble give a contribution of order lower than $0.001\,\%$, 
thus well below the statistical error. The bubble self-energy (in reciprocal space and SCHA-mode components) was explicitly computed through the formula
\begin{align}
\overset{\ssB}{\Pi}_{\mu\nu}(\bq,z)&=-\frac{1}{8N_c}\sum_{\substack{\bq_1\bq_2\\ \rho_1\rho_2}}
\frac{F(z,\omega_{\rho_1}(\bq_1),\omega_{\rho_2}(\bq_2))}{\omega_{\rho_1}(\bq_1)\omega_{\rho_2}(\bq_2)}\nonumber\\
&\overset{\sst}{D}_{\mu\rho_1\rho_2}(-\bq,\bq_1,\bq_2)\overset{\sst}{D}_{\rho_1\rho_2\nu}(-\bq_1,-\bq_2,\bq)\,,
\label{eq:app_Pi_interp}
\end{align}
where $\overset{\sst}{D}_{abc}=\overset{\sst}{\Phi}_{abc}/\sqrt{M_aM_bM_c}$.
In principle, one should sum on the $\bq$-grid commensurate with the supercell used 
to compute $\overset{\sst}{\bPhi}$ with SSCHA. However, the short-range of $\overset{\sst}{\bPhi}$ allows to interpolate it 
on a finer grid~\cite{PhysRevB.91.054304}. This accelerates, for the calculation of the self-energy, 
the convergence in the SSCHA supercell size. 

In our calculations we found sufficient to use an interpolation
grid of $12 \times 12 \times12$ (even smaller) for the static self-energy $\overset{\ssB}{\bPi}(\bq,0)$.
In Fig~.\ref{fig:static_conv} we compare the 
squared optical eigenvalue in $\Gamma$ of $\bD^{\ssF}$, for $\Hsyst$ and $\Dsyst$, 
computed with the SSCHA and using a $2\times2\times 2$ and a $3\times3\times3$ supercell. As said, $\overset{\sst}{\bPhi}$ 
has been interpolated on a $12\times 12 \times 12$ grid. The energy-force calculations were computed with DFT-PBE. 
As we can see, with the SSCHA calculation on a $2\times 2 \times 2$ supercell the result is already almost converged.

For the dynamical case, the quantity $\bPi(\bq,\omega+i0^+)$ with $\omega\neq 0$ is estimated by calculating $\bPi(\bq,\omega+i\delta)$, where $\delta$ 
is a small but finite positive value. The smaller is the value of $\delta$, the more accurate is the result, but the slower is the convergence. 
We considered $\delta=1$ meV, and an $80\times80\times80$ interpolation grid for $\overset{\sst}{\bPhi}$.

The anharmonic phonon dispersion have been computed with Eqs.~\eqref{eq:Zdef},~\eqref{eq:self_Z},~\eqref{eq:sigma_negelct_mixing_lor},~\eqref{eq:shift_width_lor}.
The self-consistent condition~\eqref{eq:self_Z} has been reached with an iterative procedure in a few steps.
In order to accelerate the SSCHA convergence on the supercell used, we interpolated on the $12 \times 12 \times 12$ also the short-range difference between the SSCHA matrices 
$\bD^{\ssS}$ (computed on the $3\times 3\times 3$ grid) and harmonic ones $\bD^{\ssz}$ on the same grid. 
Therefore, in place of the `pure' SSCHA dynamical matrices we used the matrices 
\begin{equation}
\left[\bD^{\ssS}_{3\times3\times3}-\bD^{\ssz}_{3\times3\times3}\right]_{\substack{\text{interp.}\\ 12\times12\times12}}+\bD^{\ssz}_{12\times12\times12}\,,
\end{equation}
where the subscript $m\times m\times m$ indicates the grid of the calculation.
%
\end{document}